\documentclass[twoside,english]{elsarticle}
\usepackage[T1]{fontenc}
\usepackage[latin9]{inputenc}
\usepackage{babel}
\usepackage{float}
\usepackage{graphicx}
\usepackage[unicode=true]
 {hyperref}

\makeatletter

\providecommand{\tabularnewline}{\\}

\journal{Example: Nuclear Physics B}

\makeatother

\begin{document}

\begin{frontmatter}{}

\title{Generation of a Supervised Classification Algorithm for Time-Series
Variable Stars with an Application to the LINEAR Dataset}

\author[fit]{K.B.~Johnston\fnref{fn1}\corref{cor1}\corref{cor2}}

\ead{kyjohnston2000@my.fit.edu }

\author[fit]{H.M.~Oluseyi\fnref{fn2}}

\ead{holuseyi@fit.edu}

\cortext[cor1]{Corresponding author}

\cortext[cor2]{Principal corresponding author}

\address[fit]{Florida Institute of Technology, Physic and Space Sciences Dept.,
Melbourne, Florida, 32901}
\begin{abstract}
With the advent of digital astronomy, new benefits and new problems
have been presented to the modern day astronomer. While data can be
captured in a more efficient and accurate manor using digital means,
the efficiency of data retrieval has led to an overload of scientific
data for processing and storage. This paper will focus on the construction
and application of a supervised pattern classification algorithm for
the identification of variable stars. Given the reduction of a survey
of stars into a standard feature space, the problem of using prior
patterns to identify new observed patterns can be reduced to time
tested classification methodologies and algorithms. Such supervised
methods, so called because the user trains the algorithms prior to
application using patterns with known classes or labels, provide a
means to probabilistically determine the estimated class type of new
observations. This paper will demonstrate the construction and application
of a supervised classification algorithm on variable star data. The
classifier is applied to a set of 192,744 LINEAR data points. Of the
original samples, 34,451 unique stars were classified with high confidence
(high level of probability of being the true class). \end{abstract}

\end{frontmatter}{}

\section{Introduction}

With the advent of digital astronomy, new benefits and new problems
have been presented to the modern day astronomer. While data can be
captured in a more efficient and accurate manor using digital means,
the efficiency of data retrieval has led to an overload of scientific
data for processing and storage. Where once the professional astronomer
was faced with ten to a hundred data points for a given night, the
now more common place \textquotedblleft full-sky survey\textquotedblright{}
mission results in millions of data points. This means that more stars,
in more detail, are captured per night; but increasing data capture
begets exponentially increasing data processing. No longer can the
astronomer rely on manual processing, instead the profession as a
whole has begun to adopt more advanced computational means. Database
management, digital signal processing, automated image reduction,
and statistical analysis of data have all made their way to the forefront
of tools for the modern astronomer. Astro-statistics and astro-informatics
are fields which focus on the application and development of these
tools to help aid in the processing of large scale astronomical data
resources. 

This paper will focus on one facet of this budding area, the construction
and application of a supervised pattern classification algorithm for
the identification of variable stars. Given the reduction of a survey
of stars into a standard feature space, the problem of using prior
patterns to identify new observed patterns can be reduced to time
tested classification methodologies and algorithms. Such supervised
methods, so called because the user trains the algorithms prior to
application using patterns with known (hence the supervised) classes
or labels, provides a means to probabilistically determine the estimated
class type of new observations. These methods have two large advantages
over hand-classification procedures: the rate at which new data is
processed is dependent only on the computational processing power
available and the performance of a supervised classification algorithm
is quantifiable and consistent. Thus the algorithm produces rapid,
efficient, and consistent results. 

This paper will be structured as follows. First, the data and feature
space to be implemented for training will be reviewed. Second, we
will discuss the class labels to be used and the meaning behind them.
Third, a set of classifiers (multi-layer perceptron, random forest,
k-nearest neighbor, and support vector machine) will be trained and
tested on the extracted feature space. Fourth, performance statistics
will be generated for each classifier and a comparing and contrasting
of the methods will be discussed with a \textquotedblleft champion\textquotedblright{}
classification method being selected. Fifth, the champion classification
method will be applied to the new observations to be classified. Sixth,
an anomaly detection algorithm will be generated using the so called
one-class support vector machine and will be applied to the new observations.
Lastly, based on the anomaly detection algorithm and the supervised
training algorithm a set of populations per class type will be generated.
The result will be a highly reliable set of new populations per class
type derived from the LINEAR survey.

\subsection{Prior Work}

The idea of constructing a supervised classification algorithm for
stellar classification is not unique to this paper (see \citealt{dubath2011random}
for a review), nor is the construction of a classifier for time variable
stars. Methods pursued include the construction of a detector to determine
variability (two-class classifier \citealt{barclay2011stellar}),
the design of random forests for the detection of photometric redshifts
in spectra \citet{carliles2010random}, the detection of transient
events \citet{djorgovski2012flashes} and the development of machine-assisted
discovery of astronomical parameter relationships \citet{graham2013machine}.
\citet{debosscher2009automated} explored several classification techniques
for the supervised classification of variable stars, quantitatively
comparing the performed in terms of computational speed and performance
which they took to mean accuracy. Likewise, other efforts have focused
on comparing speed and robustness of various methods (e.g. \citealt{blomme2011improved,pichara2012improved,pichara2013automatic}).
These methods span both different classifiers and different spectral
regimes, including IR surveys (\citealt{angeloni2014vvv} and \citealt{masci2014automated}),
RF surveys \citep{rebbapragada2011classification}, and optical \citep{richards2012construction}.

\section{Data}

The procedure outlined in this paper will follow the standard philosophy
for the generation of a supervised pattern classification algorithm
as professed in \citet{duda2012pattern}and\citet{hastie2004entire},
i.e. exploratory data analysis, training and testing of supervised
classifier, comparison of classifiers in terms of performance, application
of classifier. Our training data is derived from a set of three well
known variable star surveys: the ASAS survey \citep{pojmanski2005all},
the Hipparcos survey \citep{perryman1997hipparcos}, and the OGLE
dataset \citep{udalski2002optical}. Data used for this study must
meet a number of criteria:
\begin{enumerate}
\item Each star shall have differential photometric data in the u-g-r-i-z
system 
\item Each star shall have variability in the optical channel (band) that
exceeds some fixed threshold with respect to the error in amplitude
measurement
\item Each star shall have a consistent class label, should multiple surveys
address the same star
\end{enumerate}

\subsection{Sample Representation}

These requirements reduce the total training set down to 2054 datasets
with 32 unique class labels. The features extracted are based on Fourier
frequency domain coefficients\citep{deb2009light}, statistics associated
with the time domain space, and differential photometric metrics;
for more information see \citet{richards2012construction} for a table
of all 68 features with descriptions. The 32 unique class labels can
be further generalized into four main groups: eruptive, multi-star,
pulsating, and \textquotedblleft other\textquotedblright{} \citep{debosscher2009automated},
the breakdown of characterizations for the star classes follows the
following classifications:
\begin{itemize}
\item Pulsating

\begin{itemize}
\item Giants: Mira, Semireg RV, Pop. II Cepheid, Multi. Mode Cepheid
\item RR Lyrae: FO, FM, and DM
\item ``Others'' : Delta Scuti, Lambda Bootis, Beta Cephei, Slowly Pulsating
B, Gamma Doradus, SX Phe, Pulsating Be
\end{itemize}
\item Erupting: Wolf-Rayet, Chemically Peculiar, Per. Var. SG, Herbig AE/BE,
S Doradus, RCB and Classical T-Tauri
\item Multi-Star: Ellipsoidal, Beta Persei, Beta Lyrae, W Ursae Maj.
\item Other: Weak-Line T-Tauri, SARG B, SARG A, LSP, RS Cvn
\end{itemize}
The \emph{a priori} distribution of stellar classes is given in \ref{tab:Broad-Classification-of}
for the broad classes and in \ref{tab:Unique-Classification-of} for
the unique classes:

\begin{table}[h]
\caption{Broad Classification of Variable Types in the Training and Testing
Dataset \label{tab:Broad-Classification-of} }

\begin{tabular}{|c|c|c|}
\hline 
Type & Count & \% Dist\tabularnewline
\hline 
\hline 
Multi-Star & 514 & 0.25\tabularnewline
\hline 
Other & 135 & 0.07\tabularnewline
\hline 
Pulsating & 1179 & 0.57\tabularnewline
\hline 
Erupting & 226 & 0.11\tabularnewline
\hline 
\end{tabular}

\end{table}

\begin{table}[h]
\caption{Unique Classification of Variable Types in the Training and Testing
Dataset \label{tab:Unique-Classification-of}}

\begin{tabular}{|c|c|c|c|}
\hline 
Class Type & \% Dist & Class Type & \% Dist\tabularnewline
\hline 
\hline 
a. Mira & 8.0\% & m. Slowly Puls. B & 1.5\%\tabularnewline
\hline 
b1. Semireg PV & 4.9\% & n. Gamma Doradus & 1.4\%\tabularnewline
\hline 
b2. SARG A & 0.7\% & o. Pulsating Be & 2.4\%\tabularnewline
\hline 
b3. SARG B & 1.4\% & p. Per. Var. SG & 2.7\%\tabularnewline
\hline 
b4. LSP & 2.6\% & q. Chem. Peculiar & 3.7\%\tabularnewline
\hline 
c. RV Tauri & 1.2\% & r. Wolf-Rayet & 2.0\%\tabularnewline
\hline 
d. Classical Cepheid & 9.9\% & r1. RCB & 0.6\%\tabularnewline
\hline 
e. Pop. II Cepheid & 1.3\% & s1. Class. T Tauri & 0.6\%\tabularnewline
\hline 
f. Multi. Mode Cepheid & 4.8\% & s2. Weak-line T Tauri & 1.0\%\tabularnewline
\hline 
g. RR Lyrae FM & 7.2\% & s3. RS CVn & 0.8\%\tabularnewline
\hline 
h. RR Lyrae FO & 1.9\% & t. Herbig AE/BE & 1.1\%\tabularnewline
\hline 
i. RR Lyrae DM & 2.9\% & u. S Doradus & 0.3\%\tabularnewline
\hline 
j. Delta Scuti & 6.5\% & v. Ellipsoidal & 0.6\%\tabularnewline
\hline 
j1. SX Phe & 0.3\% & w. Beta Persei & 8.7\%\tabularnewline
\hline 
k. Lambda Bootis & 0.6\% & x. Beta Lyrae & 9.8\%\tabularnewline
\hline 
l. Beta Cephei & 2.7\% & y. W Ursae Maj. & 5.9\%\tabularnewline
\hline 
\end{tabular}
\end{table}

It has been shown \citep{rifkin2004defense} that how the classification
of a multi-class problem is handled can affect the performance of
the classifier; i.e. if the classifier is constructed to process all
32 unique classes as the same time, or if 32 different classifiers
(detectors) are trained individually and the results are combined
after application, or if a staged approach is best where a classifier
is trained on the four \textquotedblleft broad\textquotedblright{}
classes first then a secondary classifier is trained on the unique
class labels in each broad class \citep{debosscher2009automated}.
The \emph{a priori} distribution of classes, the number of features
to use, and the number of samples in the training set are key factors
in determining which classification procedure to use. This dependence
is often best generalized as the \textquotedblleft curse of dimensionality\textquotedblright \citep{bellman1961adaptive},
a set of problems that arise in machine learning that are tied to
attempting to quantify a signature pattern for a given class, when
the combination of a low number of training samples and high feature
dimensionality results in a sparsity of data. Increasing sparsity
results in a number of performance problems with the classifier, most
of which amount to decrease generality (over-trained classifier) and
decreased performance (low precision or high false alarm rate). Various
procedures have been developed to address the curse of dimensionality,
most often some form of dimensionality reduction technique is implemented
or a general reframing of the classification problem is performed.
For this effort, a reframing of the classification problem will be
performed to address these issues

\subsection{Feature Space}

Prior to the generation of the supervised classification algorithm,
an analysis of the training dataset is performed. This exploratory
data analysis \citep{tukey1977exploratory}, is used here to understand
the class separability prior to training, and to help the developer
gain some insight into what should be expected in terms of performance
of the final product. For example if during the course of the EDA
it is found that the classes are linearly separable in the given dimensions
using the training data, then we would expect a high performing classifier
to be possible. Likewise, initial EDA can be useful in understanding
the distribution of the classes in the given feature space answering
questions like: are the class distributions multi-dimensional Gaussian?
Do the class distributions have erratic shapes? Are they multi-modal?
Not all classifiers are good for all situations, and often an initial
qualitative EDA can help narrow down the window of which classifiers
should be investigated and provide additional intuition to the analyst.

\subsubsection{Exploratory Data Analysis}

\begin{figure}[h]
\caption{PCA applied to the ASAS+Hipp+OGLE dataset, with the broad class labels
identified and the first two principle components plotted \label{fig:PCA-applied-to}}

\includegraphics[scale=0.25]{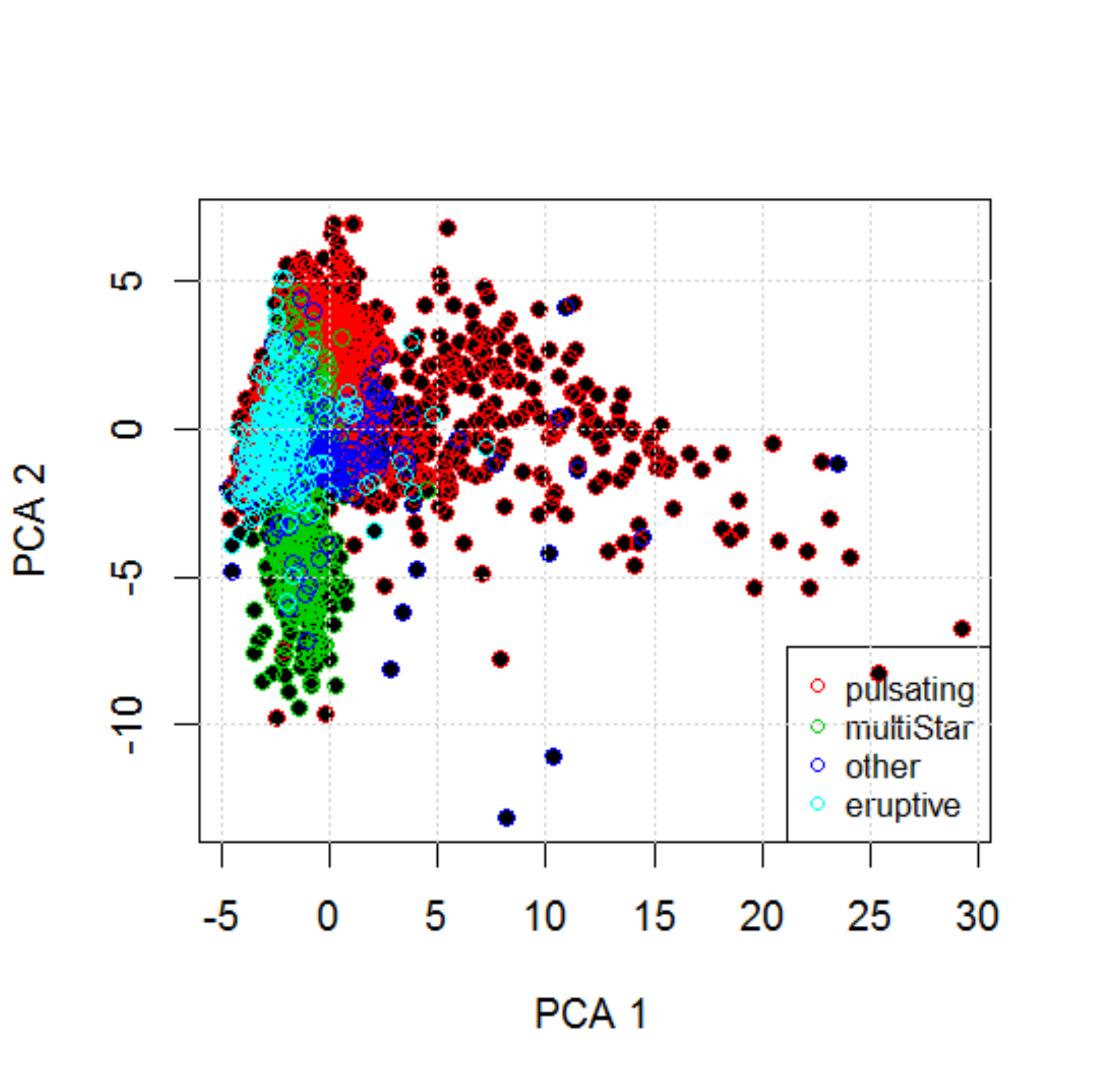}
\end{figure}

Principle Component Analysis (PCA) is one of many methods \citep{duda2012pattern},
and often the one most citied, when EDA of multi-dimensional data
is being performed. Via orthogonal transformation, PCA rotates the
feature space into a new representation where the feature dimensions
are organized such that the first dimension (the principle component)
has the largest possible variance, given the feature space. This version
of PCA is the most simple and straight-forward; there are numerous
variants, all of which attempt a similar maximization process (e.g.,
of variance, of correlation, of between group variance) but may also
employ an additional transformation (e.g., manifold mapping, using
the \textquotedblleft kernel trick\textquotedblright , etc.). Using
the broad categories defined for the variable star populations, PCA
is performed in R using the FactoMineR package \citep{le2008factominer},
and the first two components are plotted (see \ref{fig:PCA-applied-to})

\begin{figure}[h]
\caption{PCA applied to the ASAS+Hipp+OGLE dataset, only the stars classified
as \textquotedbl{}pulsating\textquotedbl{} are highlighted \label{fig:PCA-applied-to-1}}

\includegraphics[scale=0.25]{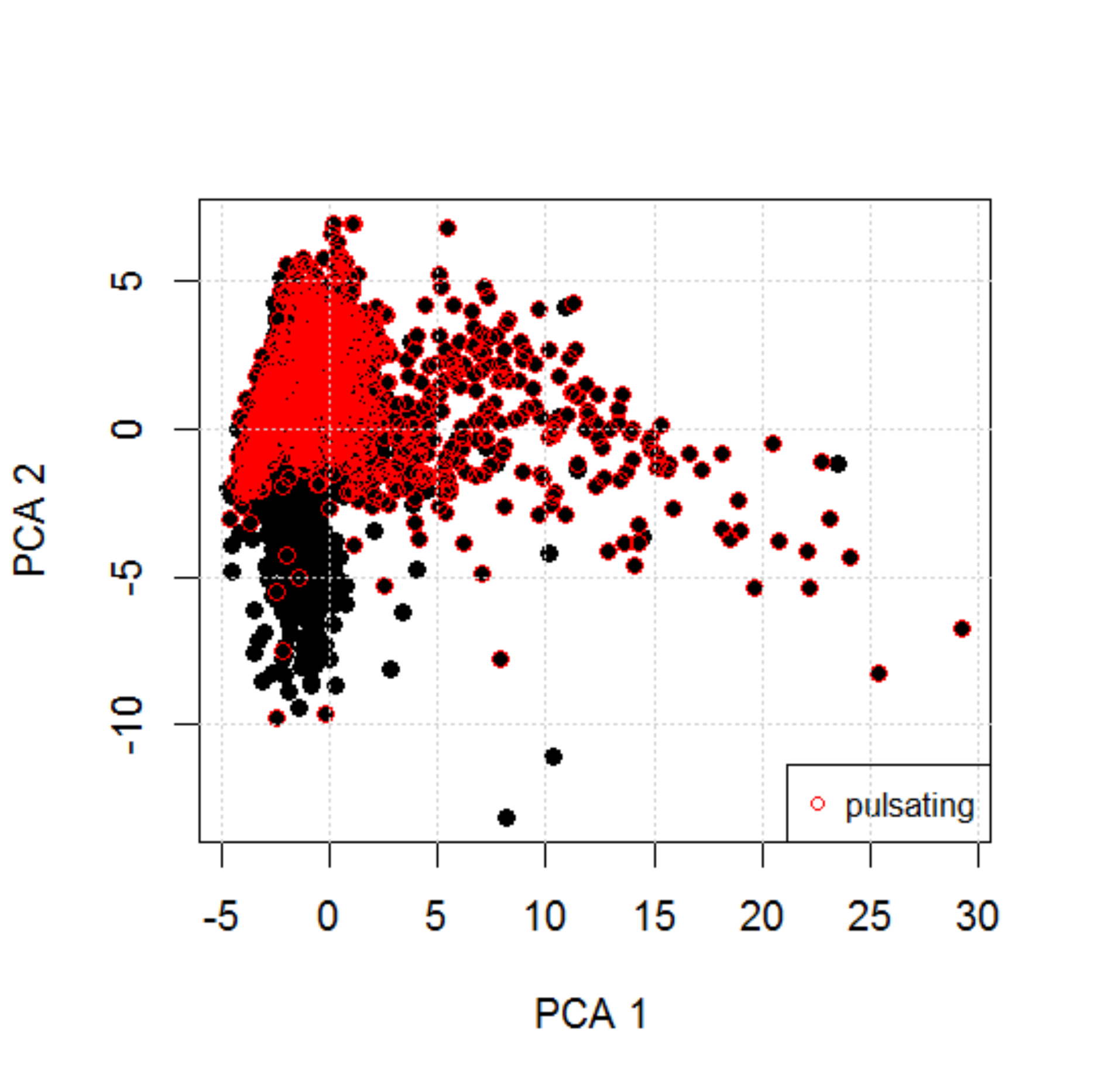}
\end{figure}

The PCA transformation is not enough to separate out the classes,
however the graphical representation of the data does provide some
additional insight about the feature space and the distribution of
classes. The eruptive and multi-star populations appear to have a
single mode in the dimensions presented in Figure 2, while the pulsating
and the \textquotedblleft other\textquotedblright{} categories appear
to be much more irregular in shape. Further analysis addressing just
the pulsating class shows that the distribution of stars with this
label is spread across the whole of the feature space (\ref{fig:PCA-applied-to-1}).

\subsection{Effectiveness of Feature Space}

In this representation of the feature space there is a significant
overlap across all classes. Even if other methods of dimensionality
reduction were implemented, for example supervised-PCA \citep{bair2006prediction},
linear separation of classes without dimensional transformation is
not possible. Application of SPCA results in the Figure 4, which is
also provided in movie form as digitally accessible media.

\begin{figure}[h]
\caption{SPCA applied to the ASAS+Hipp+OGLE dataset \label{fig:SPCA-applied-to}}

\includegraphics[scale=0.5]{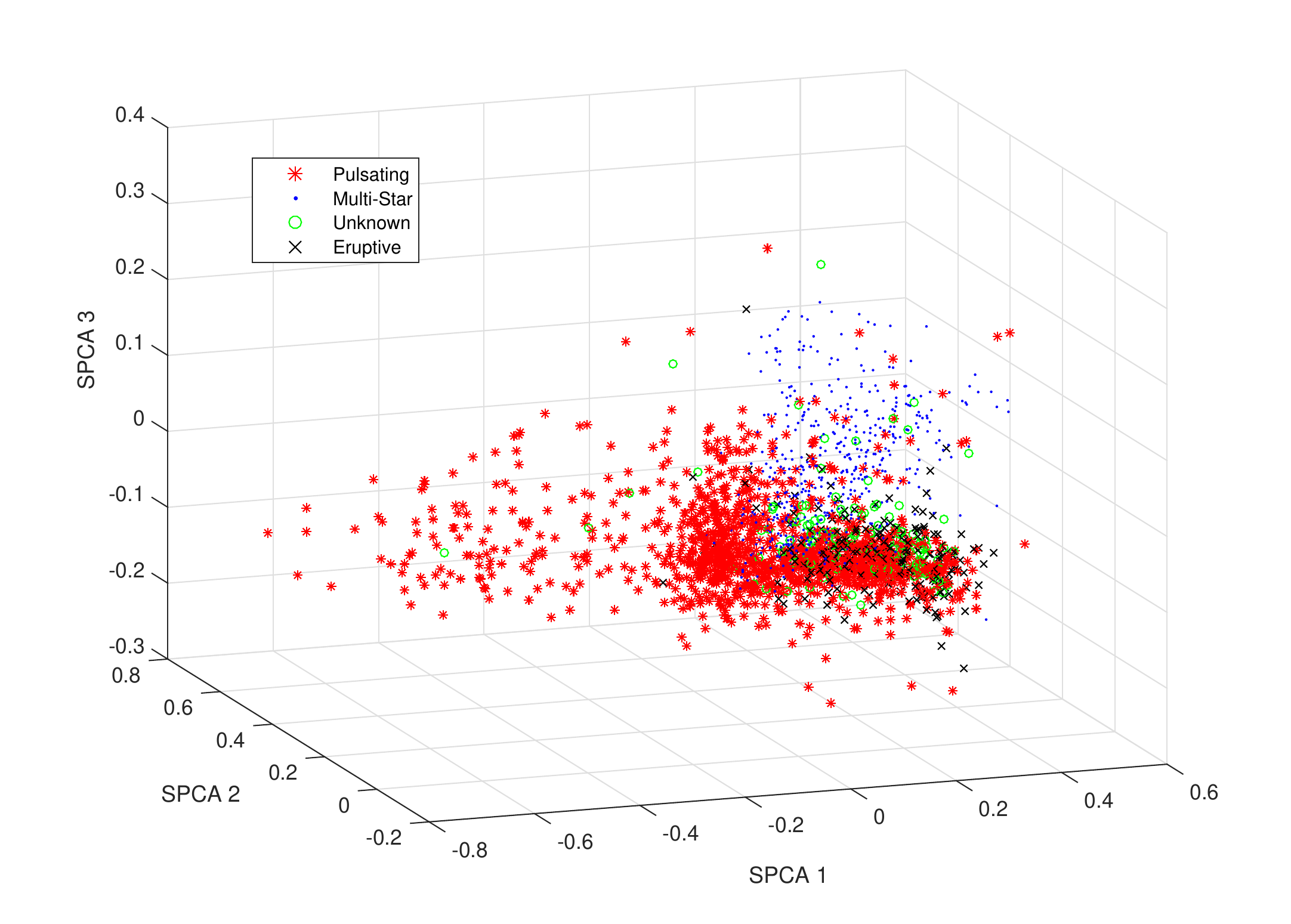}
\end{figure}

This non-Gaussian, non-linear separable class space requires further
transformation to improve separation of classes or a classifier which
performs said mapping into a space where the classes have improved
separablity. Four classifiers are briefly discussed which address
these needs.

\section{Supervised Classification Development}

All algorithms are implemented in the R language, version 3.1.2 (2014-10-31)
-- \textquotedbl{}Pumpkin Helmet\textquotedbl{}, and operations are
run on x86\_64-w64-mingw32/x64 (64-bit) platform. Four classifiers
are initially investigated: k-Nearest Neighbor (kNN), support vector
machine (SVM), random forest (RF) and multi-layer perceptron (MLP). 

The k-Nearest Neighbor algorithm implemented is based on the kNN algorithm
outlined by \citet{duda2012pattern} and \citet{altman1992introduction},
with allowance for distance measurements using both $L_{1}$ (\textquotedblleft taxi
cab\textquotedblright ) and $L_{2}$ (Euclidean distance) (see equation
1). 

$\|x\|_{p}=(|x_{1}|^{p}+|x_{2}|^{p}+...+|x_{n}|^{n})^{1/p},p=1,2,...,\infty$

The testing set is implemented to determine both optimal distance
method to be used and k value, i.e. number of nearest neighbors to
count. A number of SVM packages exist \citep{karatzoglou2005support},
the e1071 package \citep{dimitriadou2008misc} is used in this study
and was first implementation of SVM in R. It has been shown to perform
well and contains a number of additional SVM utilities beyond the
algorithm trainer that make it an ideal baseline SVM algorithm for
performance testing. SVM decisions lines are hyperplanes, linear cuts
that split the feature space into two sections; the optimal hyperplane
is the one that has the larger distance to the nearest training data
(i.e., maximum margin). Various implementations of the original algorithm
exist, including the Kernel SVM \citep{boser1992training} used here
for this study with the Gaussian (or so called radial) Kernel. KSVM
uses the so called \textquotedblleft Kernel Trick\textquotedblright{}
to project the original feature space into a higher dimension, resulting
in hyperplane decision lines that are non-linear, a beneficial functionality
should one find that the classes of interest are not linearly separable. 

The multilayer perceptron supervised learning algorithm (MLP) falls
into the family of neural network classifiers. The classifier can
be simply described as layers (stages) of perceptron (nodes), where
each perceptron performs a different transformation on the same dataset.
These perceptrons often employ simple transformation (i.e., logit,
sigmod, etc.), to go from the original input feature space, into a
set of posterior probabilities (likelihoods per estimated class label).
The construction of these layer and the transformations is beyond
the scope of this article, and for more information on neural networks,
back-propagation, error minimization and design of the classifier
the reader is invited to review such texts as \citet{rhumelhart1986parallel}.
This study makes use of the R library \textquotedblleft RSNNS,\textquotedblright{}
for the construction and analysis of the MLP classifer used; see \citet{bergmeir2012neural}.
Lastly, random forests are the conglomeration of sets of classification
and regression trees (CARTs). The CART algorithm, made popular by
\citet{breiman1984classification}, generates decisions spaces by
segmenting the feature space dimension by dimension. Given an initial
training set, the original CART is trained such that each decision
made maximally increases the purity of resulting two new populations.
Each subsequent node following either similarly divides the received
population into two new populations (with improved class purity) or
is a terminal node, where no further splits are made and instead a
class estimate is provided. 

A detailed discussion of how the CART algorithm is trained, the various
varieties of purity (or impurity) that can be used in the decision
making process, and the pruning of a constructed tree given testing
is beyond the scope of this article but are addressed in Breiman et
al. as well as other standard pattern classification text (\citealp{duda2012pattern,hastie2004entire}).
Random Forests are the conglomeration of these CART classification
algorithms, trained on variation of the same training set \citep{breiman2001random}.
This ensemble classifier constructs a set of CART algorithms, each
one trained on a reduction of the original training set (removal of
some data points in the training set), this variation results in each
CART algorithm in the set being slightly different. Given a new observed
pattern applied to the set of CART classifiers, a set of decisions
(estimated class labels) is generated. The Random Forest classifier
combines these estimated class labels to generate a unified class
estimate. This study makes use of the randomForest package in R, see
\citet{liaw2002classification}.

\subsection{Training and Testing }

The training of all four classifier types proceeds with roughly the
same procedure; following the \textquotedblleft one-vs.all\textquotedblright{}
methodology for multi-class classification, a class type of interest
is identified (either broad or unique), the original training set
is split equally into a training set and a testing set with the a
priori population distributions approximately equal to the population
distribution of the combined training set. Adjustable parameters for
each classifier are identified (RF: number of trees, kernel SVM: kernel
spread, kNN: k value and p value, MLP: number of units in the hidden
layers), and then the classifier is initially trained and tested against
the testing population. Parameters are then adjusted (and subsequent
classifiers are trained), and misclassification error is found as
a function of the parameter adjustments. Those parameters resulting
in a trained classifier with a minimal amount of error are implemented.
For each classifier, two quantifications of performance are generated:
a receiver operating curve (ROC) and a precision recall (PR) curve.
\citet{fawcett2006introduction} outlines both, and discusses the
common uses of each. Both concepts plot two performance statistics
for a given classification algorithm, given some changing threshold
value, which will for this study be a critical probability that the
posterior probability of the class of interest (the target stellar
variable) is compared against. These curves can be generated when
the classifier is cast as a \textquotedblleft two-class\textquotedblright{}
problem, where one of the classes is the target (class of interest)
while the other is not. 

For any two-class classifier the metrics highlighted here can be generated
and are a function of the decision space selected by the analyst.
Frequently the acceptance threshold, i.e. the \textquotedblleft hypothesized
class\textquotedblright{} must have a posterior probability greater
than some $\lambda$, is selected based on the errors of the classifier.
Many generic classification algorithms are designed such that the
false positive (fp) rate and 1 \textendash{} true positive (tp) rate
are both minimized. Often this practice is ideal; however the problem
faced in the instances addressed in this article require additional
considerations. We note two points: 
\begin{enumerate}
\item When addressing the \textquotedblleft unique\textquotedblright{} class
types, there are a number of stellar variable populations which are
relatively much smaller than others. This so called \textquotedblleft class
imbalance\textquotedblright{} has been shown \citep{fawcett2006introduction}
to cause problems with performance analysis if not handled correctly.
Some classification algorithms adjust for this imbalance, but often
additional considerations must be made, specifically when reporting
performance metrics.
\item Minimization of both errors, or minimum-error-rate classification,
is often based on the zero-one loss function. In this case, it is
assumed that the cost of a false positive (said it was, when it really
was not) is the same as a false negative (said it wasn\textquoteright t,
when it really was). If the goal of this study is to produce a classifier
that is able to classify new stars from very large surveys, some of
which are millions of stars big, the cost of returning a large number
of false alarms is much higher than the cost of missing some stars
in some classes. Especially when class separation is small (if not
non-existent), if the application of the classifier results in significant
false alarms the inundation of an analyst with bad decisions will
likely result in a general distrust of the classifier algorithm.
\end{enumerate}
The ROC curve expresses the adjustment of the errors as a function
of the decision criterion. Likewise, the PR curve expresses the adjustment
of precision (the percentage of true positives out of all decisions
made) and recall (true positive rate) as a function of the decision
criterion. By sliding along the ROC or PR curve, we can change the
performance of the classifier. Note that increasing the true positive
rate causes an increase in the false positive rate as well (and vice
versa). Often a common practice is to fix \citep{scharf1991statistical}
one of the metrics, false positive rate, of all classifiers used.
For example, minimum requirements might be a false positive rate of
5\%, if the ROC curve was resulting from a classifier designed; we
would expect a 20\% true positive rate. 

Similar to the ROC curve, the PR curve demonstrates how performance
varies between precision and recall for a given value of the threshold.
It is apparent that the PR and ROC curves are related \citep{davis2006relationship},
both have true-positive rate as an axis (TP Rate and Recall are equivalent),
both are functions of the threshold used in the determination of estimated
class for a new patter (discrimination), both are based on the confusion
matrix and the associated performance metrics. Thus fixing the false
alarm rate, not only fixes the true positive rate but also the precision
of the classifier. If the interest was a general comparison of classifiers,
instead of selecting a specific performance level, \citet{fawcett2006introduction}
suggests that the computation of Area-Under-the-Curve quantifies either
the PR and ROC curve into a single \textquotedblleft performance\textquotedblright{}
estimate that represents the classifier as a whole. The ROC-AUC of
a classifier is equivalent to the probability that the classifier
will rank a randomly chosen positive instance higher than a randomly
chosen negative instance. The PR-AUC of a classifier is roughly the
mean precision of the classifier. Both ROC and PR curves should be
considered when evaluating a classifier \citep{davis2006relationship},
especially when class imbalances exist. For this study, the best performing
classifier will be the one that maximizes both the ROC-AUC and the
PR-AUC. Likewise, when the final performance of the classifier is
proposed, false positive rate and precision will be reported and used
to make assumption about the decisions made by the classification
algorithm.

\subsection{Performance Analysis - Supervised Classification}

Based on the foundation of performance analysis methods, ROC and PR
curves, and AUC discussed, the study analyzes classification algorithms
applied to both the broad and unique (individual) class labels.

\subsubsection{Broad Classes - Random Forest}

Initially an attempt was made to adjust both the number of trees (ntree)
and the number of variables randomly sampled as candidates at each
split (mtry). It was found that for the training datasets, that neither
resulted in a major difference in performance when adjusted. Based
on \citet{breiman1984classification} recommendation, mtry was set
to $\sqrt{M}$, where $M$ is the number of features. The parameters
ntree was set to 100, based on the work performed by \citet{debosscher2009automated}.
Classifiers were then generated based on the training sample, and
the testing set was used to generate the ROC and PR AUC for each one-vs-all
classifier. The associated curves are given in Appendix A, the resulting
AUC estimates are in \ref{tab:ROC/PR-AUC-Estimates}:

\begin{table}[h]
\caption{ROC/PR AUC Estimates based on training and testing for the Random
Forest Classifier. \label{tab:ROC/PR-AUC-Estimates}}

\begin{tabular}{|c|c|c|c|c|}
\hline 
AUC & Pulsating & Eruptive & Multi-Star & Other\tabularnewline
\hline 
\hline 
ROC & 0.971 & 0.959 & 0.992 & 0.961\tabularnewline
\hline 
PR & 0.979 & 0.788 & 0.986 & 0.800\tabularnewline
\hline 
\end{tabular}

\end{table}

\subsubsection{Broad Classes - Kernel SVM}

Instead of using the \textquotedblleft hard\textquotedblright{} class
estimates, common with SVM usage, the \textquotedblleft soft\textquotedblright{}
estimates i.e. posterior probabilities are used. This allows for the
thresholding necessary to construct the PR and ROC curves. Kernel
spreads of 0.001, 0.01, and 0.1 were tested (set as the variable gamma
in R), the associated PR and ROC curve are given in Appendix A. It
was found that 0.1 was optimal for the feature space (using Gaussian
Kernels). The associated curves are given in Appendix A, the resulting
AUC estimates are in \ref{tab:ROC/PR-AUC-Estimates-1}:

\begin{table}[h]
\caption{ROC/PR AUC Estimates based on training and testing for the Kernel
SVM Classifier. \label{tab:ROC/PR-AUC-Estimates-1}}

\begin{tabular}{|c|c|c|c|c|}
\hline 
AUC & Pulsating & Eruptive & Multi-Star & Other\tabularnewline
\hline 
\hline 
ROC & 0.938 & 0.903 & 0.979 & 0.954\tabularnewline
\hline 
PR & 0.952 & 0.617 & 0.968 & 0.694\tabularnewline
\hline 
\end{tabular}
\end{table}

\subsubsection{Broad Classes - k-NN}

It was found, that for the training set used, that increasing performance
was gained with increasing values of k (number of nearest neighbors).
Gains in performance were limiting after k = 4, and a value of k =
10 was selected to train with. The value of the polynomial defined
in the generation of distance (via $L^{p}$-norm) was varied between
1 and 3, with decreasing performance found for p > 3. The associated
PR and ROC curves were generated for values of p < 4. The associated
curves are given in Appendix A, the resulting AUC estimates for p
< 3 are in \ref{tab:ROC/PR-AUC-Estimates-2}:

\begin{table}[h]
\caption{ROC/PR AUC Estimates based on training and testing for the k-NN Classifier.\label{tab:ROC/PR-AUC-Estimates-2}}

\begin{tabular}{|c|c|c|c|c|c|}
\hline 
 & AUC & Pulsating & Eruptive & Multi-Star & Other\tabularnewline
\hline 
\hline 
p-1 & ROC & 0.919 & 0.847 & 0.980 & 0.928\tabularnewline
\hline 
 & PR & 0.931 & 0.480 & 0.959 & 0.597\tabularnewline
\hline 
p-2 & ROC & 0.901 & 0.802 & 0.967 & 0.877\tabularnewline
\hline 
 & PR & 0.918 & 0.368 & 0.931 & 0.519\tabularnewline
\hline 
\end{tabular}
\end{table}

\subsubsection{Broad Classes - MLP}

There are two variables associated with MLP algorithm training: the
number of units in the hidden layers (size) and the number of parameters
for the learning function to use (learnParam). It was found that for
the dataset: The learnParam function had little effect on the performance
of the classifier, and it was taken to be 0.1 for implementation here.
The variable size did have an effect, an initial study of values between
4 and 18 demonstrated that the best performance occurred between 4
and 8. PR and ROC curves for these values were generated and are in
Appendix A, the resulting AUC estimates for values 4,6 and 8 are in
\ref{tab:ROC/PR-AUC-Estimates-3}:

\begin{table}[h]
\caption{ROC/PR AUC Estimates based on training and testing for the MLP Classifier.
\label{tab:ROC/PR-AUC-Estimates-3}}

\begin{tabular}{|c|c|c|c|c|c|}
\hline 
 & AUC & Pulsating & Eruptive & Multi-Star & Other\tabularnewline
\hline 
\hline 
 & ROC & 0.928 & 0.694 & 0.914 & 0.585\tabularnewline
\hline 
 & PR & 0.916 & 0.120 & 0.869 & 0.183\tabularnewline
\hline 
 & ROC & 0.933 & 0.751 & 0.888 & 0.473\tabularnewline
\hline 
 & PR & 0.909 & 0.139 & 0.797 & 0.123\tabularnewline
\hline 
 & ROC & 0.920 & 0.706 & 0.914 & 0.529\tabularnewline
\hline 
 & PR & 0.903 & 0.175 & 0.854 & 0.159\tabularnewline
\hline 
\end{tabular}
\end{table}

\subsubsection{Unique Classes}

\begin{figure}[h]
\caption{ROC vs. PR AUC Plot for Multi-Layer Perceptron Classifier, generic
class types (Eruptive, Giants, Cephids, RRlyr, Other Pulsing, Multi-star,
and \textquotedblleft other\textquotedblright ) are colored. The line
y = x is plotted for reference (dashed line)\label{fig:ROC-vs.-PR}}

\includegraphics[scale=0.25]{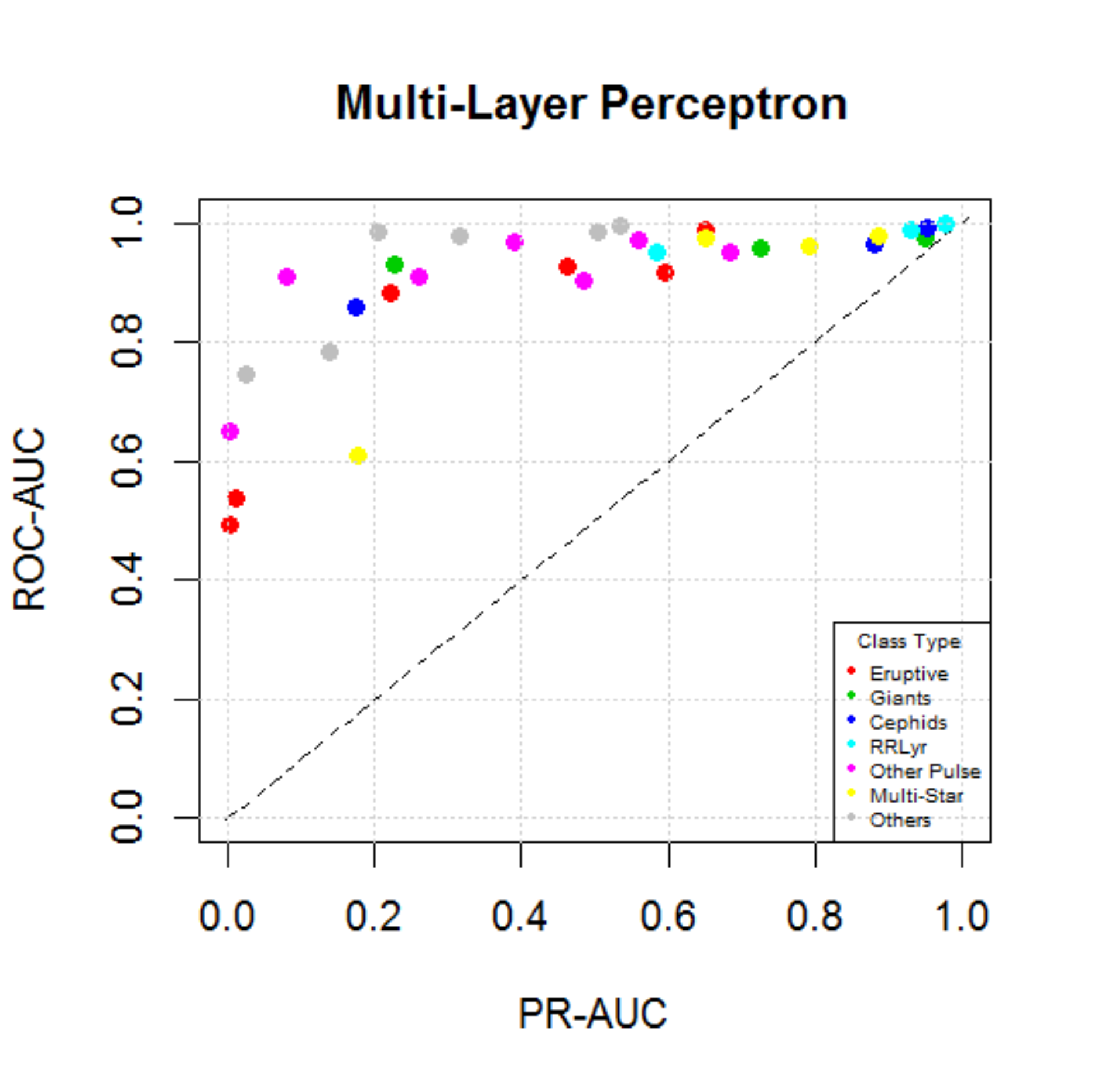}
\end{figure}

Analysis of the broad classes provided insight into the potential
of a staged classifier. The performance of the broad classification
algorithms does not suggest that the supervised variable star classification
problem would be benefited by a staged design. The RF classifier performed
best across all broad classes and against all other classifiers, but
still had significant error; had the broad classes perfectly separated
further analysis into the staged design would have been warranted.
Instead, two-class classifier designed based on the unique classes
are explored. Similar to the broad classification methodology, the
training sample is separated into a training set and a testing data
set for each unique class type for training in a two-class classifier.
Again, the testing data is used to minimize the misclassification
error, and find optimal parameters for each of the classifiers. Each
classifier is then optimal for the particular class of interest. With
the change of design, a change of performance analysis is also necessary.
With nearly ten times the number of classes, a comparison of ROC and
PR curves per classifier type and per class type requires a methodology
that allows the information plotted on a single plot (direct comparison).
Keeping with the discussion outlined by \citet{davis2006relationship},
we plot ROC vs. PR for each classifier (\ref{fig:ROC-vs.-PR}as an
example). 

The set of these performance analysis graphs are in Appendix B. A
comparison of the general performance of each classifier can be derived
from the generation of AUC for each of the performance curves. Here,
the quantification of general performance for a classifier is given
as either mean precision across all class types or via non-parametric
analysis of the AUC. The non-parametric analysis used is compiled
as follows: for each class of stars, the average performance across
classifiers is found, if for a classifier the performance is greater
than the mean, an assignment of +1 is given, else -1. Over all classes,
the summation of assignments is taken and given in \ref{tab:Performance-Analysis-of}.

\begin{table}[H]
\caption{Performance Analysis of Individual Classifiers \label{tab:Performance-Analysis-of}}

\begin{tabular}{|c|c|c|c|c|}
\hline 
 & \multicolumn{2}{c|}{ROC-AUC} & \multicolumn{2}{c|}{PR-AUC}\tabularnewline
\hline 
 & Mean & Non-Para. & Mean & Non-Para.\tabularnewline
\hline 
\hline 
KNN-Poly-1 & 0.884 & 2 & 0.530 & 8\tabularnewline
\hline 
SVM & 0.905 & -4 & 0.407 & -26\tabularnewline
\hline 
MLP & 0.894 & 2 & 0.470 & -8\tabularnewline
\hline 
RF & 0.948 & 22 & 0.595 & 14\tabularnewline
\hline 
\end{tabular}
\end{table}

It is apparent that the RF classifier out-performs the other three
classification algorithms, using both the mean of precision as well
as a non-parametric comparison of the AUC statistics. The plot comparing
ROC-AUC and PR-AUC for the Random Forest classifier is presented in
\ref{fig:Random-Forest,-Individual}.

\begin{figure}[h]
\caption{Random Forest, Individual Classification, Performance Analysis \label{fig:Random-Forest,-Individual}}

\begin{tabular}{|c|c|}
\hline 
\includegraphics[scale=0.25]{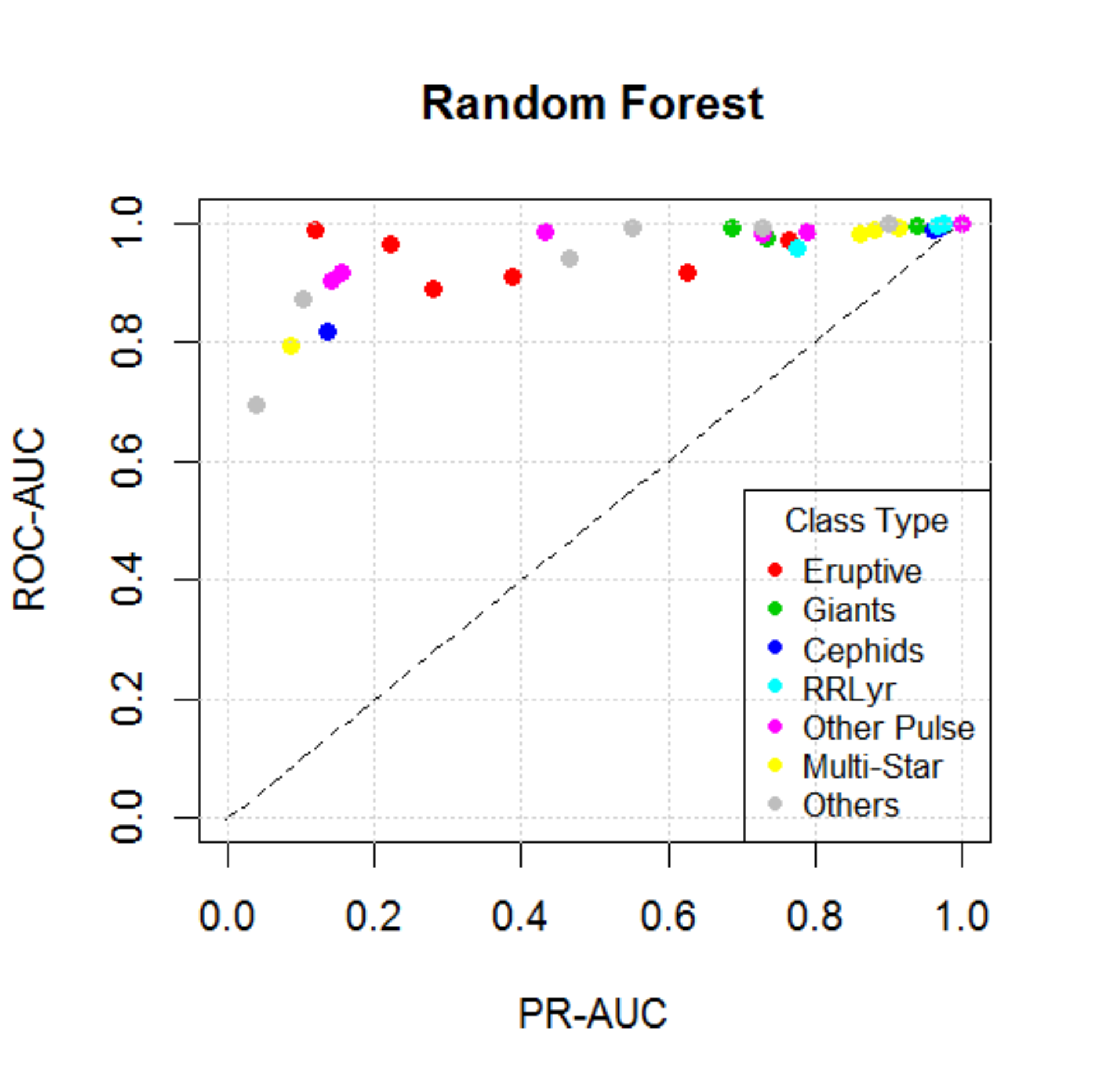}(a) & \includegraphics[scale=0.25]{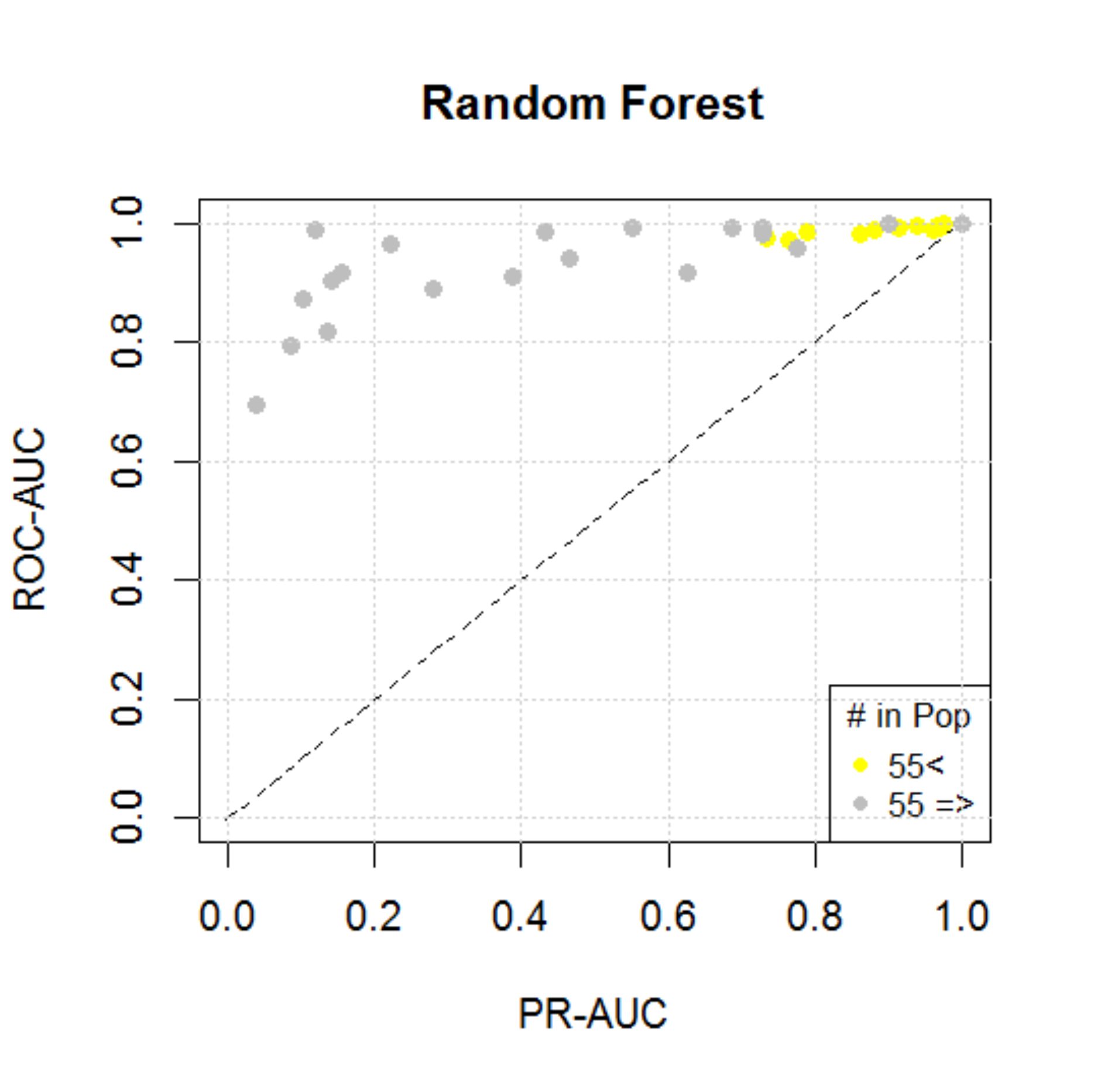}(b)\tabularnewline
\hline 
\end{tabular}
\end{figure}

Based on \ref{fig:Random-Forest,-Individual}, it is observed that
star populations of similar class types do not necessarily cluster
together. Additionally it is apparent that the original size of the
population in the training set, while having some effect on the ROC-AUC,
has a major effect on the resulting PR-AUC. Figure 11.b demonstrates
that for those classes with an initial population of 55 (empirically
guessed value), that the precision is expected to be greater than
70\%. Surprisingly though for classes with an initial population of
55 or less, the limits of precision are less predictable and in fact
appear to be random with respect to class of interest training size.
Thus, without further training data or feature space improvements,
the performance statistics graphed in \ref{fig:Random-Forest,-Individual}
are the statistics that will be used as part of the application of
the classifier to the LINEAR dataset.

\subsection{Performance Analysis - Anomaly Detection}

\begin{figure}[h]
\caption{Fraction of Anomalous Points Found in the Training Dataset as a Function
of the Gaussian Kernel Spread Used in the Kernel-SVM \label{fig:Fraction-of-Anomalous}}

\includegraphics[scale=0.25]{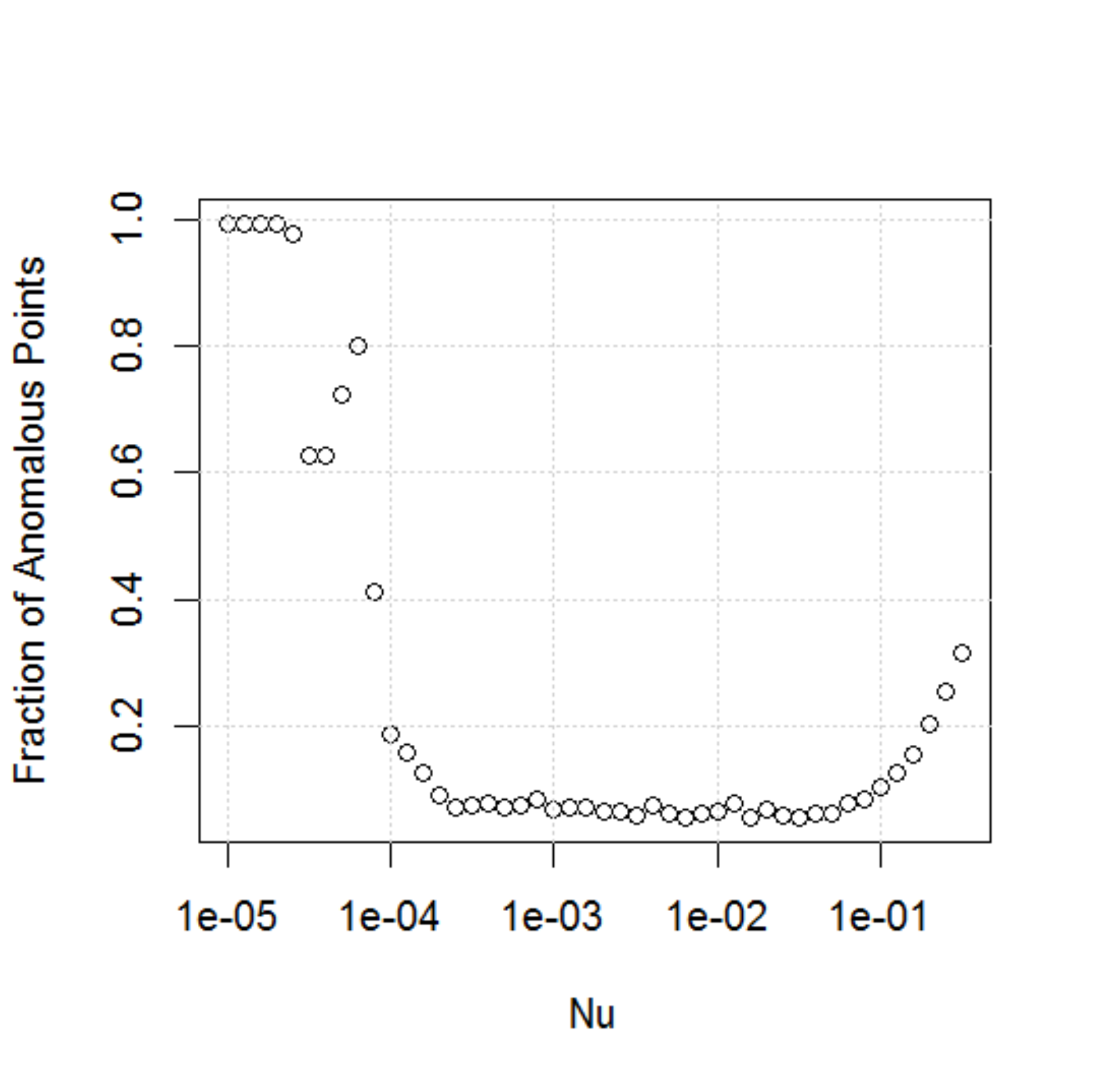}

\end{figure}

In addition to the pattern classification algorithm outlined, the
procedure outlined here includes the construction of a One-Class Support
Vector Machine (OC-SVN) for use as an anomaly detector. The pattern
classification algorithms presented and compared as part of this analysis,
partition the entire decision space. For the random forest, kNN, MLP
and SVM two-class classifier algorithms, there is no consideration
for deviations of patterns beyond the training set observed, i.e.
absolute distance from population centers. All of the algorithms investigated
consider relative distances, i.e. is the new pattern P closer to the
class center of B or A? Thus, despite that an anomalous pattern is
observed by a new survey, the classifier will attempt to estimate
a label for the observed star based on the labels it knows. To address
this concern, a one-class support vector machine is implemented as
an anomaly detection algorithm. Lee and Scott (2007) describe the
design and construction of such an algorithm. Similar to the Kernel-SVM
discussed prior, the original dimensionality is expanded using the
Kernel trick (Gaussian Kernels) allowing complex regions to be more
accurately modeled. For the OC-SVM, the training data labels are adjusted
such that all entered data is of class type one (+1). A single input
pattern at the origin point is artificially set as class type two
(-1). The result is the \textquotedblleft lassoing\textquotedblright{}
or dynamic encompassing of \textquotedblleft known\textquotedblright{}
data patterns. The lasso boundary represents the division between
known (previously observed) regions of feature space and unknown (not-previously
observed) regions. New patterns observed with feature vectors occurring
in this unknown region are considered anomalies or patterns without
support, and the estimated labels returned from the supervised classification
algorithms should be questioned, despite the associated posterior
probability of the label estimate \citep{scholkopf2001estimating}.
The construction of the OC-SVM to be applied as part of this analysis
starts with the generation of two datasets (training and testing)
from the ASAS + Hipp + OGLE training data. The initial training set
is provided to the OC-SVM \citep{lee2007one} algorithm which generates
the decision space (lasso). This decision space is tested against
the training data set; and the fraction of points declared to be anomalous
is plotted against the spread of the Kernel used in the OC-SVM (\ref{fig:Fraction-of-Anomalous}).

\begin{figure}[h]
\caption{Plot of OC-SVM Results Applied to Training Data Only \label{fig:Plot-of-OC-SVM}}

\includegraphics[scale=0.25]{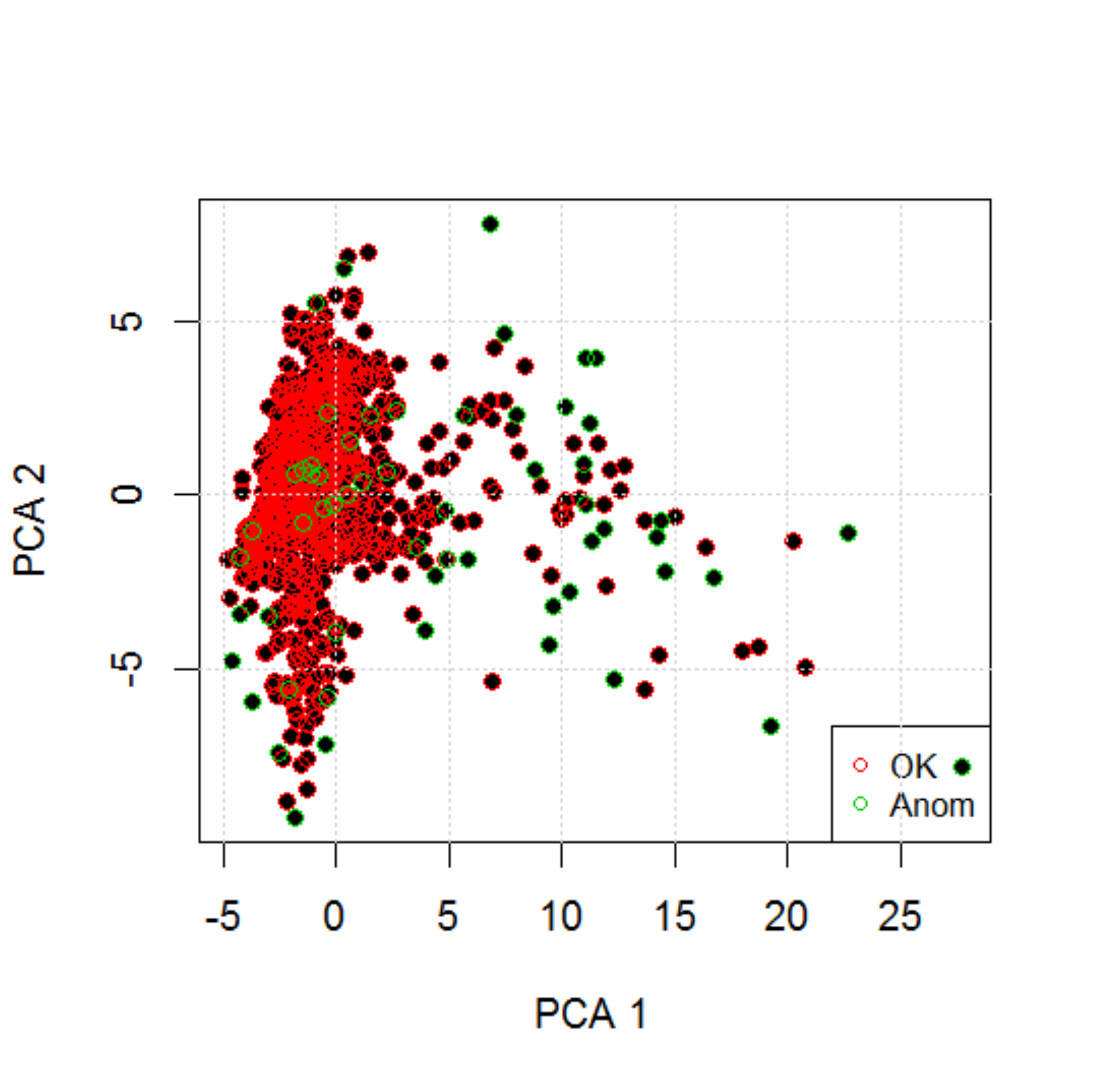}

\end{figure}

Because of the hyper-dimensionality, the OC-SVM algorithm is unable
to perfectly encapsulate the training data; however a minimization
can be found and estimated. The first two principle components of
the training data feature space are plotted for visual inspection
(\ref{fig:Plot-of-OC-SVM}), highlighting those points that were called
\textquotedblleft anomalous\textquotedblright{} based on a nu value
(kernel spread) of 0.001. Less than 5\% of the points are referred
to as anomalies (\textasciitilde{}falsely).

\begin{figure}[h]
\caption{OC-SVM testing of the Testing data \label{fig:OC-SVM-testing-of}}

\includegraphics[scale=0.25]{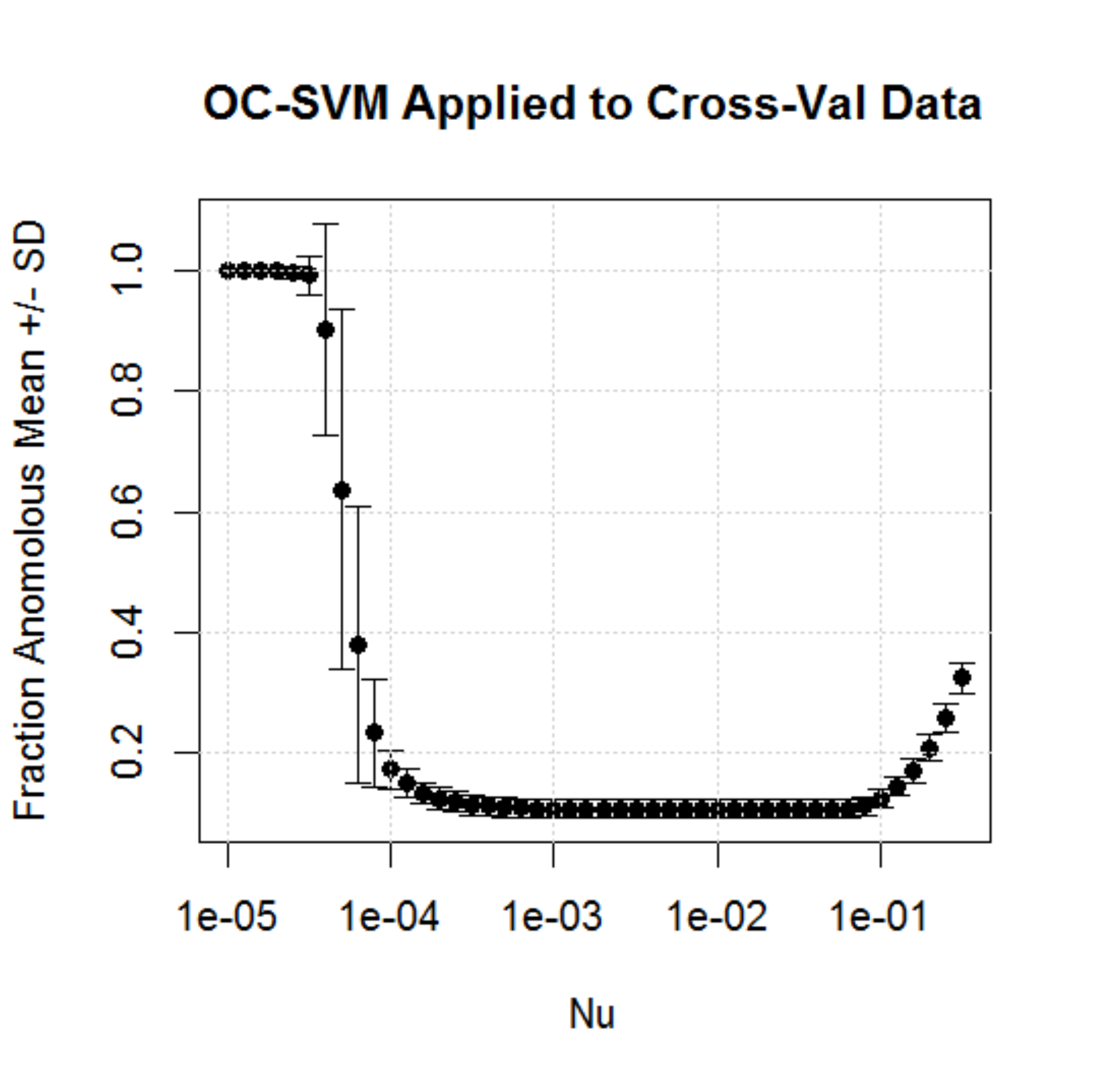}

\end{figure}

Further testing is performed on the anomaly space, using the second
dataset generated. As both datasets originate from the same parent
population, the OC-SVM algorithm parameter (nu) is tuned to a value
that maximally accepts the testing points (\ref{fig:OC-SVM-testing-of}).

The minimum fraction was found at a nu of 0.03. The OC-SVM was applied
to the LINEAR dataset with the optimal kernel spread. All 192,744
datasets were processed, with 58,312 (False, or \textquotedblleft anomalous\textquotedblright )
and 134,432 (True, or \textquotedblleft expected\textquotedblright )
decision made, i.e. \textasciitilde{}30\% of the LINEAR dataset is
considered anomalous based on the ASAS+HIPP+OGLE training dataset
feature space.

\section{Application of Supervised Classifer to LINEAR Dataset}

\subsection{Analysis and Results}

For application to the LINEAR dataset, a RF classifier is constructed
based on the training set discussed prior. The classifiers are designed
using the one-vs.-all methodology, i.e. each stellar class has its
own detector (i.e. overlap in estimated class labels is possible),
therefore 32 individual two-class classifiers (detectors) are generated.
The individual classification method (one-vs.-all) allows for each
given star to have multiple estimated labels (e.g. multiple detectors
returning a positive result for the same observation). The one-vs.-all
methodology also allows the training step of the classification to
be more sensitive to stars who might have been under-represented in
the training sample, improving the performance of the detector overall.
Based on the testing performance results (ROC and PR curves) presented
for the individual classifiers, the critical statistic used for the
RF decision process was tuned such that a 0.5\% false positive rate
is expected when applied to the LINEAR dataset. In addition to the
RF classifier, an OC-SVM anomaly detection algorithm was trained,
an used to determine if samples from the LINEAR dataset are anomalous
with respect to the ASAS+OGLE+HIPP dataset. Applying the RF classifier(s)
and the OC-SVM algorithm to the LINEAR dataset the following was found
using a threshold setting corresponding to a false alarm rate of 0.5\%
(see ROC curve analysis). Given an initial set of LINEAR data (192,744
samples), the following table was constructed based on the results
of the application of the isolated one-vs.-all RF classifiers only: 

\begin{table}[h]
\caption{Initial results from the application of the RF classifier(s) \label{tab:Initial-results-from}}

\begin{tabular}{|c|c|c|c|}
\hline 
Class Type & Est. Pop & Class Type & Est. Pop\tabularnewline
\hline 
\hline 
a. Mira & 3256 & m. Slowly Puls. B & 2\tabularnewline
\hline 
b1. Semireg PV & 7 & n. Gamma Doradus & 2268\tabularnewline
\hline 
b2. SARG A & 4291 & o. Pulsating Be & 14746\tabularnewline
\hline 
b3. SARG B & 30 & p. Per. Var. SG & 284\tabularnewline
\hline 
b4. LSP & 10 & q. Chem. Peculiar & 10\tabularnewline
\hline 
c. RV Tauri & 5642 & r. Wolf-Rayet & 3970\tabularnewline
\hline 
d. Classical Cepheid & 31 & r1. RCB & 1253\tabularnewline
\hline 
e. Pop. II Cepheid & 326 & s1. Class. T Tauri & 17505\tabularnewline
\hline 
f. Multi. Mode Cepheid & 556 & s2. Weak-line T Tauri & 4945\tabularnewline
\hline 
g. RR Lyrae FM & 13470 & s3. RS CVn & 40512\tabularnewline
\hline 
h. RR Lyrae FO & 1276 & t. Herbig AE/BE & 1358\tabularnewline
\hline 
i. RR Lyrae DM & 9800 & u. S Doradus & 2185\tabularnewline
\hline 
j. Delta Scuti & 493 & v. Ellipsoidal & 132\tabularnewline
\hline 
j1. SX Phe & 9118 & w. Beta Persei & 481\tabularnewline
\hline 
k. Lambda Bootis & 69 & x. Beta Lyrae & 2\tabularnewline
\hline 
l. Beta Cephei & 2378 & y. W Ursae Maj. & 1365\tabularnewline
\hline 
\end{tabular}
\end{table}

103628 stars were not classified (\textasciitilde{}54\%) and of those
11619 were considered \textquotedblleft Anomalous\textquotedblright .
57848 stars were classified only once (30\%) and of those 23397 were
considered \textquotedblleft Anomalous\textquotedblright . 31268 stars
were classified with multiple labels (\textasciitilde{}16\%) and of
those 23296 were considered \textquotedblleft Anomalous\textquotedblright .
The set of stars that were both classified once and did not have anomalous
patterns (34,451), are broken down by class type in \ref{tab:Initial-results-from-1}.

\begin{table}[h]
\caption{Initial results from the application of the RF classifier(s) and the
OC-SVM anomaly detection algorithm, classes that are major returned
classes (>1\% of the total return set) are in bold \label{tab:Initial-results-from-1}}

\begin{tabular}{|c|c|c|c|c|c|}
\hline 
Class Type & Est. Pop & \% Total & Class Type & Est. Pop & \% Total\tabularnewline
\hline 
\hline 
a. Mira & 15 & 0.04\% & m. Slowly Puls. B & 2 & 0.002\%\tabularnewline
\hline 
b1. Semireg PV & 1 & 0.002\% & \textbf{n. Gamma Doradus} & \textbf{2268} & \textbf{3.8\%}\tabularnewline
\hline 
\textbf{b2. SARG A} & \textbf{1362} & \textbf{4.0\%} & o. Pulsating Be & 14746 & 0.61\%\tabularnewline
\hline 
b3. SARG B & 0 & 0\% & p. Per. Var. SG & 284 & 0.26\%\tabularnewline
\hline 
b4. LSP & 1 & 0.002\% & q. Chem. Peculiar & 10 & 0\%\tabularnewline
\hline 
\textbf{c. RV Tauri} & \textbf{538} & \textbf{1.6\%} & \textbf{r. Wolf-Rayet} & \textbf{3970} & \textbf{6.2\%}\tabularnewline
\hline 
d. Classical Cepheid & 2 & 0.006\% & r1. RCB & 1253 & 0.01\%\tabularnewline
\hline 
e. Pop. II Cepheid & 50 & 0.15\% & \textbf{s1. Class. T Tauri} & \textbf{17505} & \textbf{5.4\%}\tabularnewline
\hline 
f. Multi. Mode Cepheid & 286 & 0.83\% & \textbf{s2. Weak-line T Tauri} & \textbf{4945} & \textbf{3.3\%}\tabularnewline
\hline 
\textbf{g. RR Lyrae FM} & \textbf{2794} & \textbf{8.1\%} & \textbf{s3. RS CVn} & \textbf{40512} & \textbf{46.6\%}\tabularnewline
\hline 
\textbf{h. RR Lyrae FO} & \textbf{710} & \textbf{2.1\%} & t. Herbig AE/BE & 1358 & 0.33\%\tabularnewline
\hline 
\textbf{i. RR Lyrae DM} & \textbf{2350} & \textbf{6.8\%} & \textbf{u. S Doradus} & \textbf{2185} & \textbf{1.7\%}\tabularnewline
\hline 
j. Delta Scuti & 8 & 0.02\% & v. Ellipsoidal & 132 & 0.08\%\tabularnewline
\hline 
\textbf{j1. SX Phe} & \textbf{1624} & \textbf{4.7\%} & w. Beta Persei & 481 & 0.42\%\tabularnewline
\hline 
k. Lambda Bootis & 1 & 0.002\% & x. Beta Lyrae & 2 & 0.006\%\tabularnewline
\hline 
l. Beta Cephei & 25 & 0.07\% & \textbf{y. W Ursae Maj.} & \textbf{1365} & \textbf{3.1\%}\tabularnewline
\hline 
\end{tabular}
\end{table}

The listing of individual discovered populations are provided digitally
via request (\href{http://kyjohnst2000@my.fit.edu}{to the author}).
Two classes were not detected confidently out of the LINEAR dataset:
SARG B and Chemically Peculiar. This does not mean that these stars
are not contained in the LINEAR dataset. Similarly, those stars that
were not classified are not necessarily in a \textquotedblleft new\textquotedblright{}
class of stars. There are a number of possibilities why these stars
were not found in the survey including: 
\begin{enumerate}
\item Poor separation between the class of interest (for a given detector)
and other stars. Poor separation could result in either the posterior
probability not being high enough to detect the star, or more likely
the star being classified as two different types at the same time
\item Poor initial quantification of the signature class pattern in the
training set feature space. If the training sample representing a
given class type spanned only a segment of the signature class pattern
region, the potential for an under-sampled or poorly bounding feature
space exists. Furthermore, application of the anomaly detection algorithm,
or any of the pattern classification algorithms, would result in decision-lines
lassoing the under-sampled feature space, cutting through the \textquotedblleft true
class pattern region\textquotedblright . New observations of that
class type, if they occurred outside of the original under-sampled
space, would likely be flagged by the anomaly detection algorithm
or as a different class. 
\end{enumerate}
Thus, those stars positively classified by the set of detectors used
represent the set of LINEAR observations that have patterns that are
consistent with those observed in the training set. As part of the
testing process we estimate both a false alarm rate (FAR) of 0.5\%
across all classes and a precision rate from the PR curve. Then, each
one-vs.-all detector will have a different precision rate, since the
FAR is fixed. The precision rate estimates based on testing are given
in \ref{tab:Precision-Rate-Estimates}. An adjusted estimate of \textquotedblleft true\textquotedblright{}
returned population sizes can be estimated by considering the precision
rate, i.e., if 15 Mira stars were detected, and the Mira detector
had a precision of \textasciitilde{}94\% percent, then potentially
1 of those detections is a false positive.

\begin{table}[h]
\caption{Precision Rate Estimates Per Class Type (in fractions), Bolded Classes
are those with Precisions < 80\% \label{tab:Precision-Rate-Estimates}}

\begin{tabular}{|c|c|c|c|}
\hline 
Class Type & Precision & Class Type & Precision\tabularnewline
\hline 
\hline 
a. Mira & 0.94 & m. Slowly Puls. B & 0.91\tabularnewline
\hline 
b1. Semireg PV & 0.97 & n. Gamma Doradus & 0.88\tabularnewline
\hline 
\textbf{b2. SARG A} & \textbf{0.76} & o. Pulsating Be & 0.91\tabularnewline
\hline 
b3. SARG B & 0.94 & p. Per. Var. SG & 0.94\tabularnewline
\hline 
b4. LSP & 0.91 & q. Chem. Peculiar & 0.94\tabularnewline
\hline 
c. RV Tauri & 0.86 & r. Wolf-Rayet & 0.91\tabularnewline
\hline 
d. Classical Cepheid & 0.94 & \textbf{r1. RCB} & \textbf{0.73}\tabularnewline
\hline 
e. Pop. II Cepheid & 0.87 & \textbf{s1. Class. T Tauri} & \textbf{0.75}\tabularnewline
\hline 
f. Multi. Mode Cepheid & 0.87 & \textbf{s2. Weak-line T Tauri} & \textbf{0.75}\tabularnewline
\hline 
g. RR Lyrae FM & 0.91 & \textbf{s3. RS CVn} & \textbf{0.74}\tabularnewline
\hline 
h. RR Lyrae FO & 0.88 & t. Herbig AE/BE & 0.86\tabularnewline
\hline 
\textbf{i. RR Lyrae DM} & \textbf{0.73} & \textbf{u. S Doradus} & \textbf{0.67}\tabularnewline
\hline 
j. Delta Scuti & 0.95 & \textbf{v. Ellipsoidal} & \textbf{0.78}\tabularnewline
\hline 
\textbf{j1. SX Phe} & \textbf{0.73} & w. Beta Persei & 0.97\tabularnewline
\hline 
k. Lambda Bootis & 0.91 & x. Beta Lyrae & 0.97\tabularnewline
\hline 
l. Beta Cephei & 0.91 & y. W Ursae Maj. & 0.91\tabularnewline
\hline 
\end{tabular}
\end{table}

\begin{table}[h]
\caption{True Recovery Estimates based on Precision Estimates and the Number
of Stars Detected by the RF classifier \label{tab:True-Recovery-Estimates}}

\begin{tabular}{|c|c|c|c|}
\hline 
Class Type & Est. Pop & Class Type & Est. Pop\tabularnewline
\hline 
\hline 
a. Mira & 14 & m. Slowly Puls. B & 0\tabularnewline
\hline 
b1. Semireg PV & 0 & n. Gamma Doradus & 1159\tabularnewline
\hline 
b2. SARG A & 1035 & o. Pulsating Be & 192\tabularnewline
\hline 
b3. SARG B & 0 & p. Per. Var. SG & 85\tabularnewline
\hline 
b4. LSP & 0 & q. Chem. Peculiar & 0\tabularnewline
\hline 
c. RV Tauri & 462 & r. Wolf-Rayet & 1939\tabularnewline
\hline 
d. Classical Cepheid & 1 & r1. RCB & 2\tabularnewline
\hline 
e. Pop. II Cepheid & 43 & s1. Class. T Tauri & 1383\tabularnewline
\hline 
f. Multi. Mode Cepheid & 248 & s2. Weak-line T Tauri & 843\tabularnewline
\hline 
g. RR Lyrae FM & 2542 & s3. RS CVn & 11850\tabularnewline
\hline 
h. RR Lyrae FO & 624 & t. Herbig AE/BE & 96\tabularnewline
\hline 
i. RR Lyrae DM & 1715 & u. S Doradus & 387\tabularnewline
\hline 
j. Delta Scuti & 7 & v. Ellipsoidal & 21\tabularnewline
\hline 
j1. SX Phe & 1185 & w. Beta Persei & 138\tabularnewline
\hline 
k. Lambda Bootis & 0 & x. Beta Lyrae & 1\tabularnewline
\hline 
l. Beta Cephei & 22 & y. W Ursae Maj. & 986\tabularnewline
\hline 
\end{tabular}

\end{table}

\section{Conclusions}

With the onset of large scale stellar surveys, methods such as the
supervised classifier are becoming more and more necessary. The volume
of data returned by these surveys has exceeded the amount that can
be hand processed in any reasonable amount of time. Data science,
statistics, digital signal processing, and other exploratory data
analysis methods are then necessary to produce at a minimum the actionable
information demonstrated in this paper. This paper outlines the application
of one of these tools, supervised classification algorithms, to be
used in the identification of stellar variables. Variable stars provide
an opportunity to observe not only differential photometric features,
but also single (or multi-band) time-domain feature, providing a large
feature space with which stellar classes can be separated. Time domain
features can include descriptive statistics associated with the time
domain variability, or the transformation of the time domain feature
into a basis representation that is constant over the life-time of
observations of the star. Accurate handling of these features can
provide separablity in classes, allowing machine operations to rapidly
categorize new observations. 

This paper has demonstrated the construction and application of a
supervised classification algorithm on variable star data. Such an
algorithm, will process observed stellar features and produce quantitative
estimates of stellar class label. Using a hand-process (verified)
dataset derived from the ASAS, OGLE, and Hipparcos survey, an initial
training and testing set was derived. The trained one-vs.-all algorithms
were optimized using the testing data via minimization of the misclassification
rate. From application of the trained algorithm to the testing data,
performance estimates can be quantified for each one-vs.-all algorithm.
The Random Forest supervised classification algorithm was found to
be superior for the feature space and class space operated in. Similarly,
a one-class support vector machine was trained in a similar manor,
and designed as an anomaly detector. 

With the classifier and anomaly detection algorithm constructed, both
were applied to a set of 192744 LINEAR data points. Of the original
samples, Setting the threshold of the RF classifier using a false
alarm rate of 0.5\%, 34,451 unique stars were classified only once
in the one-vs.-all scheme and were not identified by the anomaly detection
algorithm. The total population is partitioned into the individual
stellar variable classes; each subset of LINEAR ID corresponding to
the matched patterns is stored in a separate file and accessible to
the reader. While less than 18\% of the LINEAR data was classified,
the class labels estimated have a high level of probability of being
the true class based on the performance statistics generated for the
classifier, and the threshold applied to the classification process. 

Further improvement in both the initial training dataset is necessary,
if the requirements of the supervised classification algorithm are
to be met (100\% classification of new data). Larger training data,
with more representation (support) is needed to improve the class
space representation used by the classifier, and reduce the size of
the \textquotedblleft anomalous\textquotedblright{} decision region.
Specifically, additional example of the under-sampled variable stars,
enough to perform k-fold cross-validation would yield improved performance
and increased generality of the classifier. An improved feature space
could also benefit the process, if new features were found to provide
additional linear separation for certain classes. However, additional
dimensionality without reduction of superfluous features is warned
against as it may only worsen the performance issues of the classifier.
Instead, investigation into the points found to be anomalous in under-sampled
classes, and determination if they are indeed of the class reported
by the classifier designed here would be of benefit, as this points
would serve to not only bolster the number of training points used
in the algorithm, but they would also increase the size (and support)
of the individual class spaces. Implementation of these concepts,
with a mindfulness of the changing performance of the supervised classification
algorithm, could result in performance improvements across the class
space. 

\bibliographystyle{plainnat}
\bibliography{SupervisedRef}

\appendix

\section{Broad Class Performance Results}

\begin{figure}[h]

\caption{Random Forest, mtry = 8, ntree = 100, (a) Pulsating, (b) Erupting,
(c) Multi-Star, (d) Other \label{fig:Random-Forest,-mtry}}

\begin{tabular}{|c|c|}
\hline 
\includegraphics[scale=0.25]{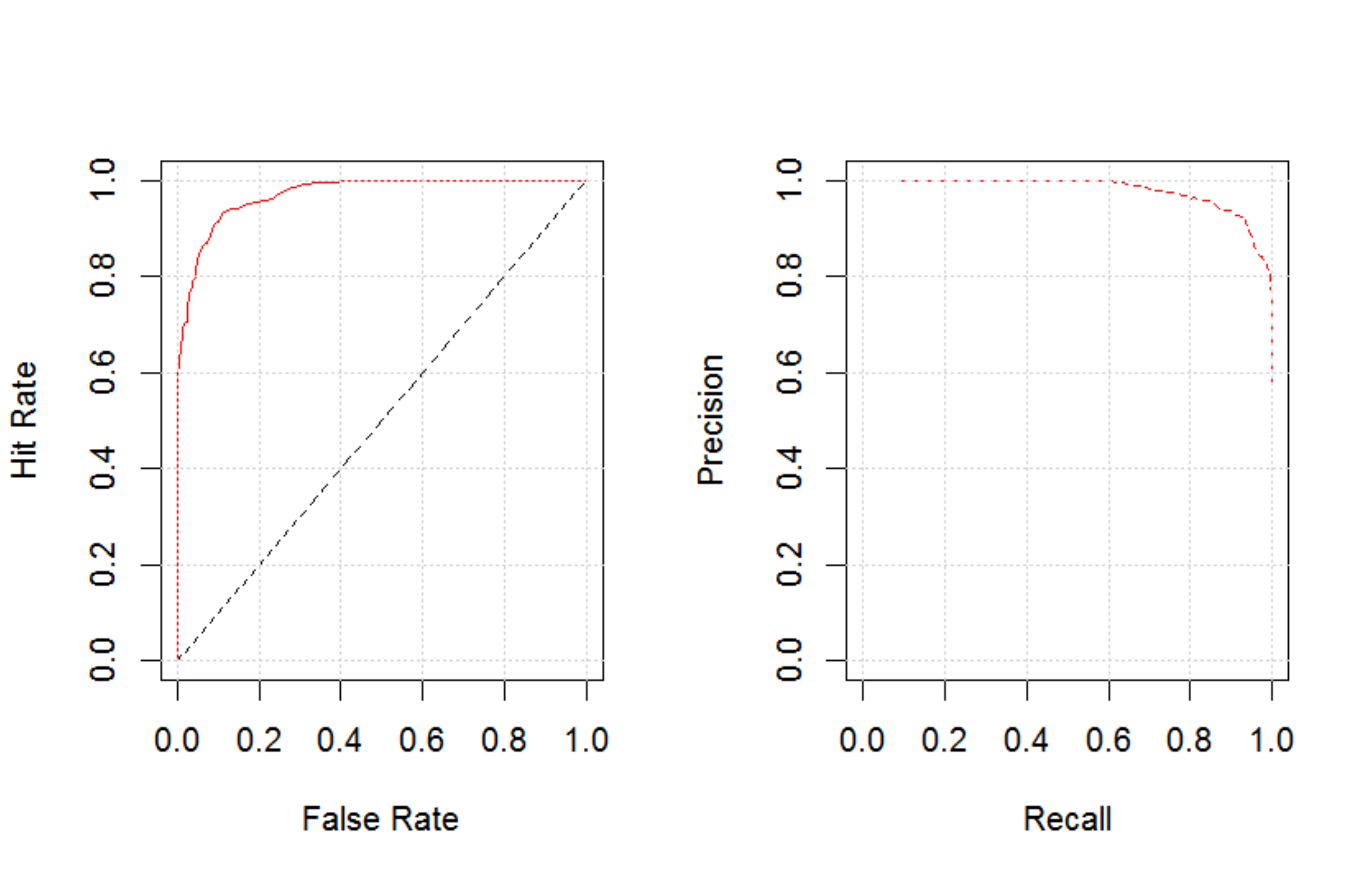}(a) & \includegraphics[scale=0.25]{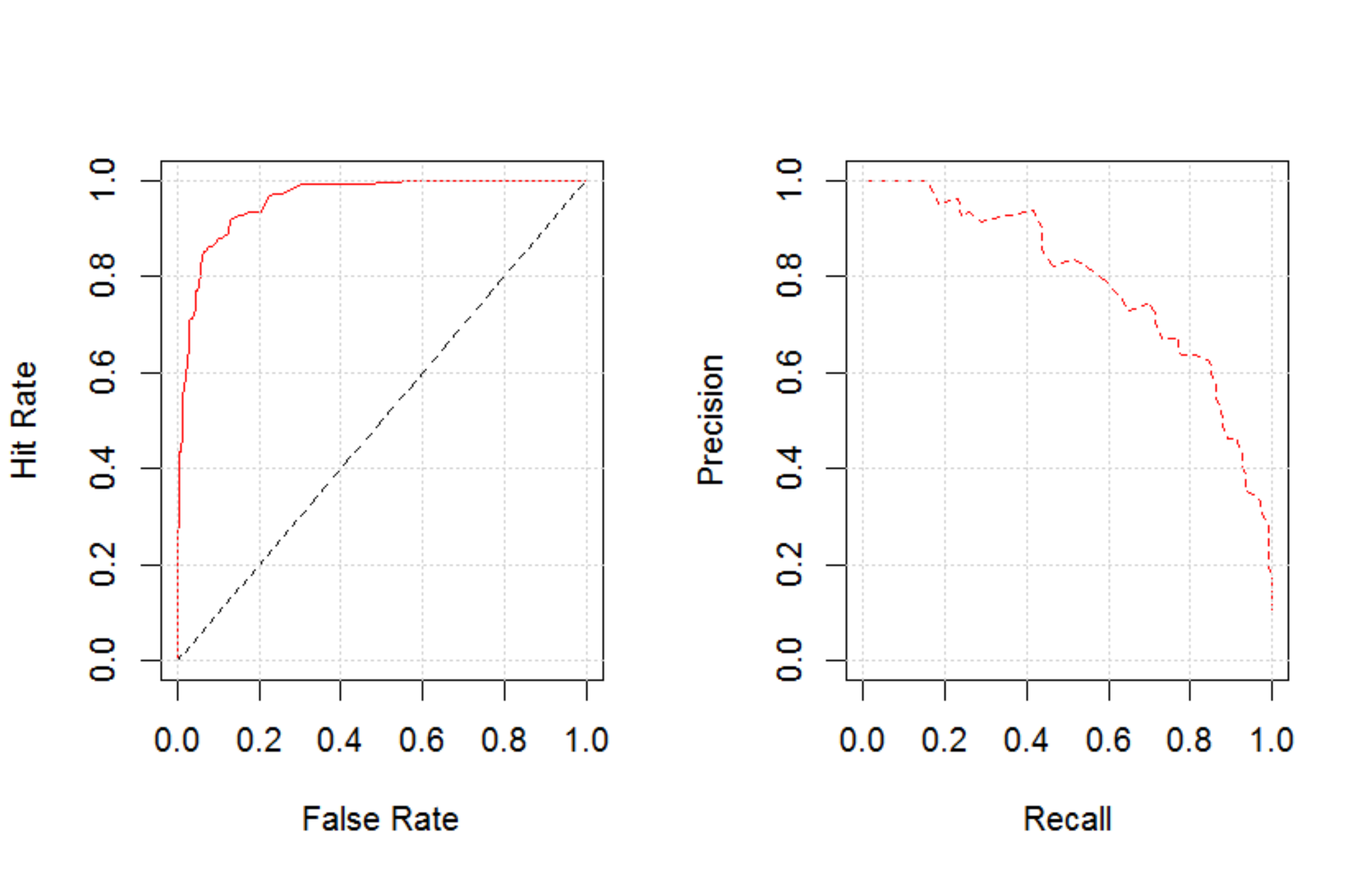}(b)\tabularnewline
\hline 
\hline 
\includegraphics[scale=0.25]{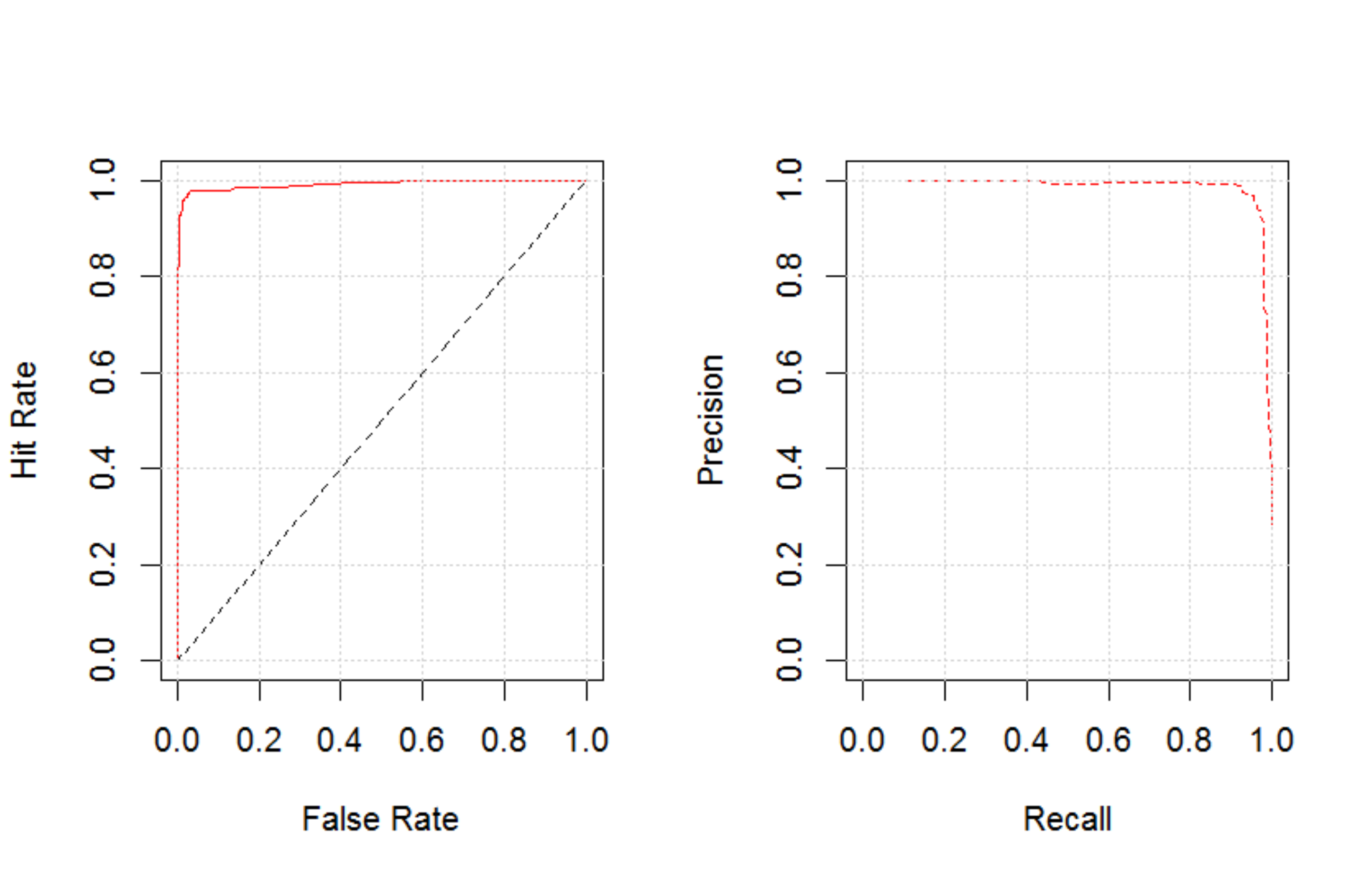}(c) & \includegraphics[scale=0.25]{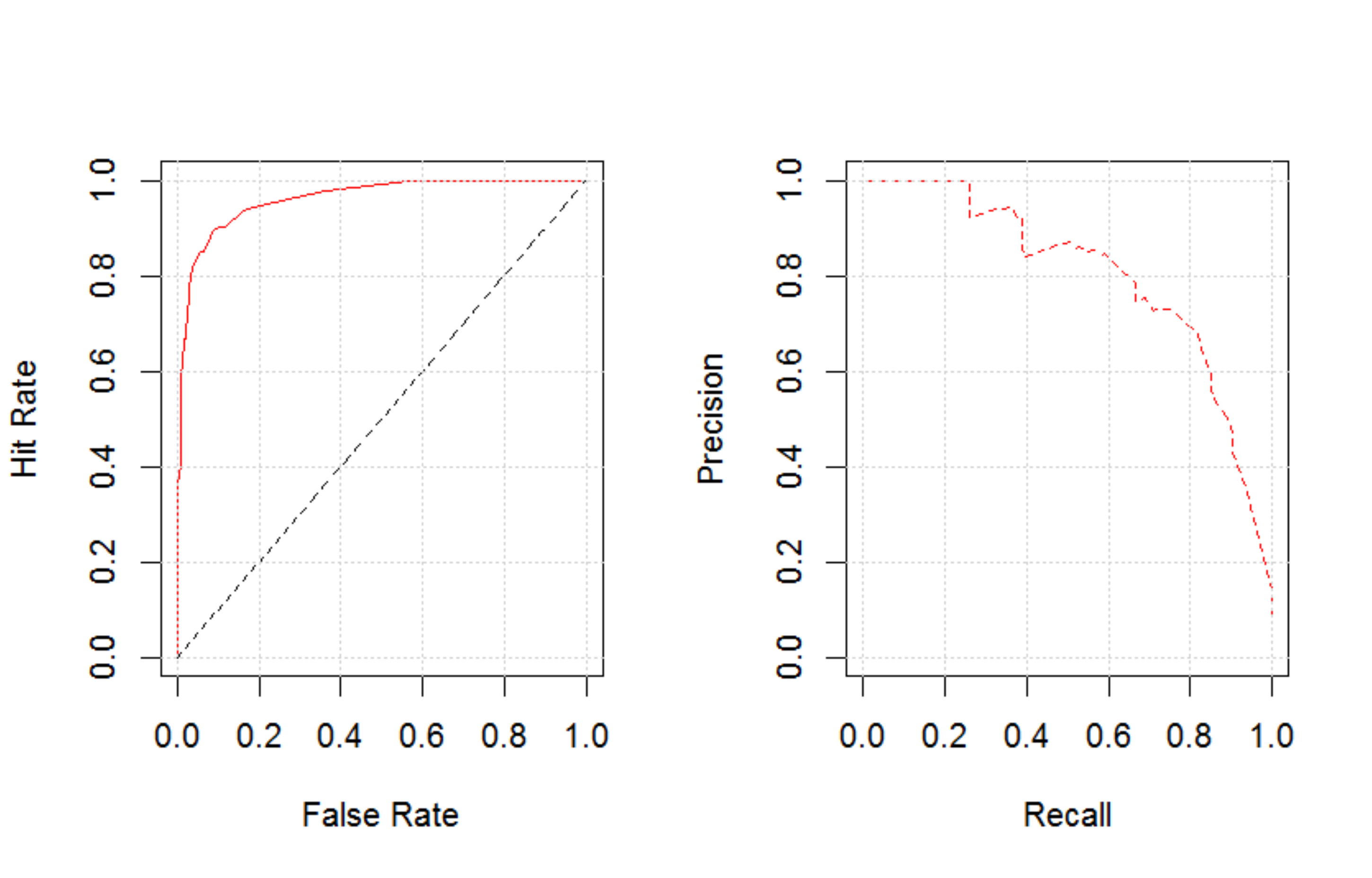}(d)\tabularnewline
\hline 
\end{tabular}
\end{figure}

\begin{figure}[h]

\caption{SVM, (a) Pulsating, (b) Erupting, (c) Multi-Star, (d) Other \label{fig:SVM,-(a)-Pulsating,}}

\begin{tabular}{|c|c|}
\hline 
\includegraphics[scale=0.25]{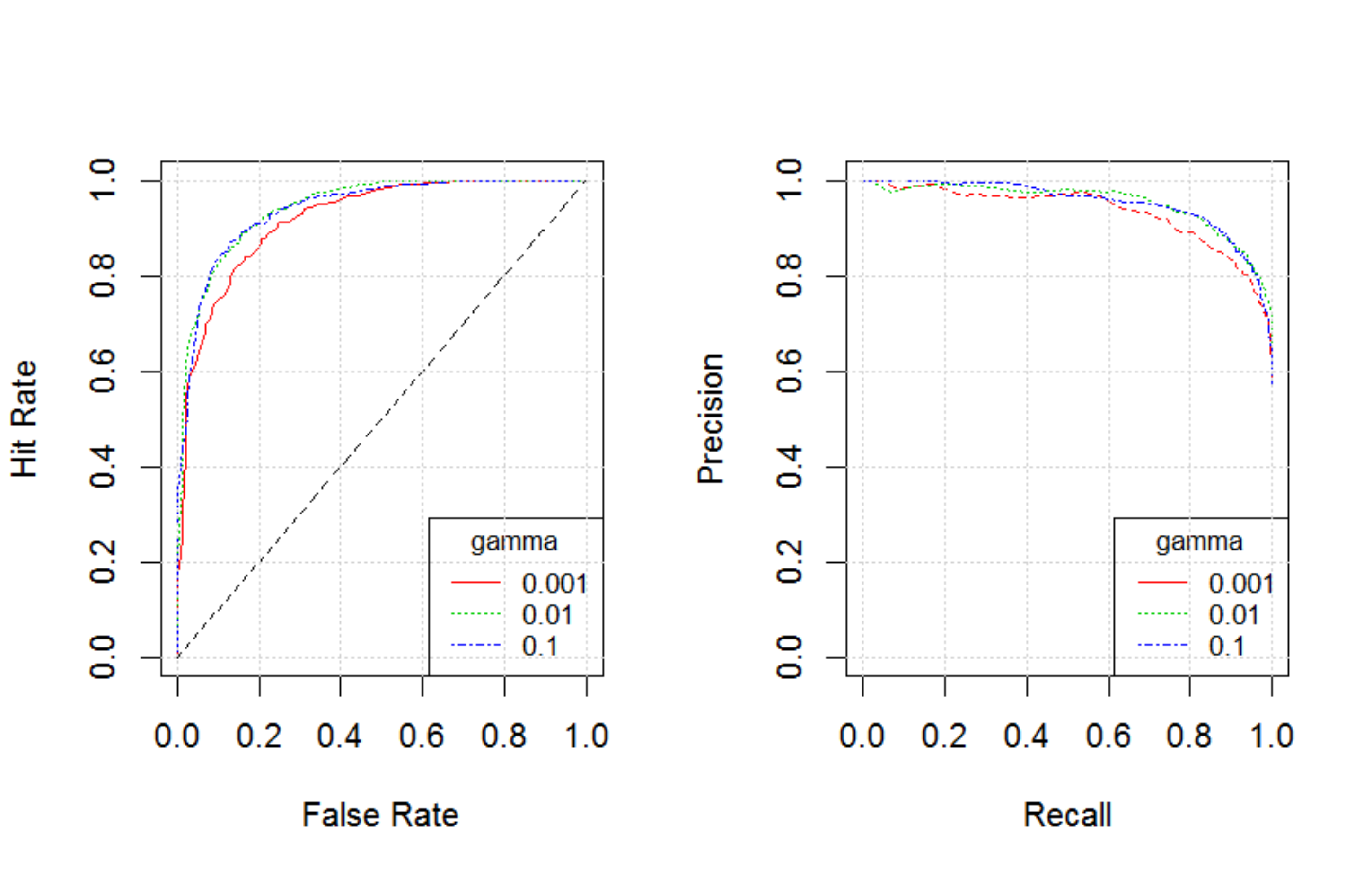}(a) & \includegraphics[scale=0.25]{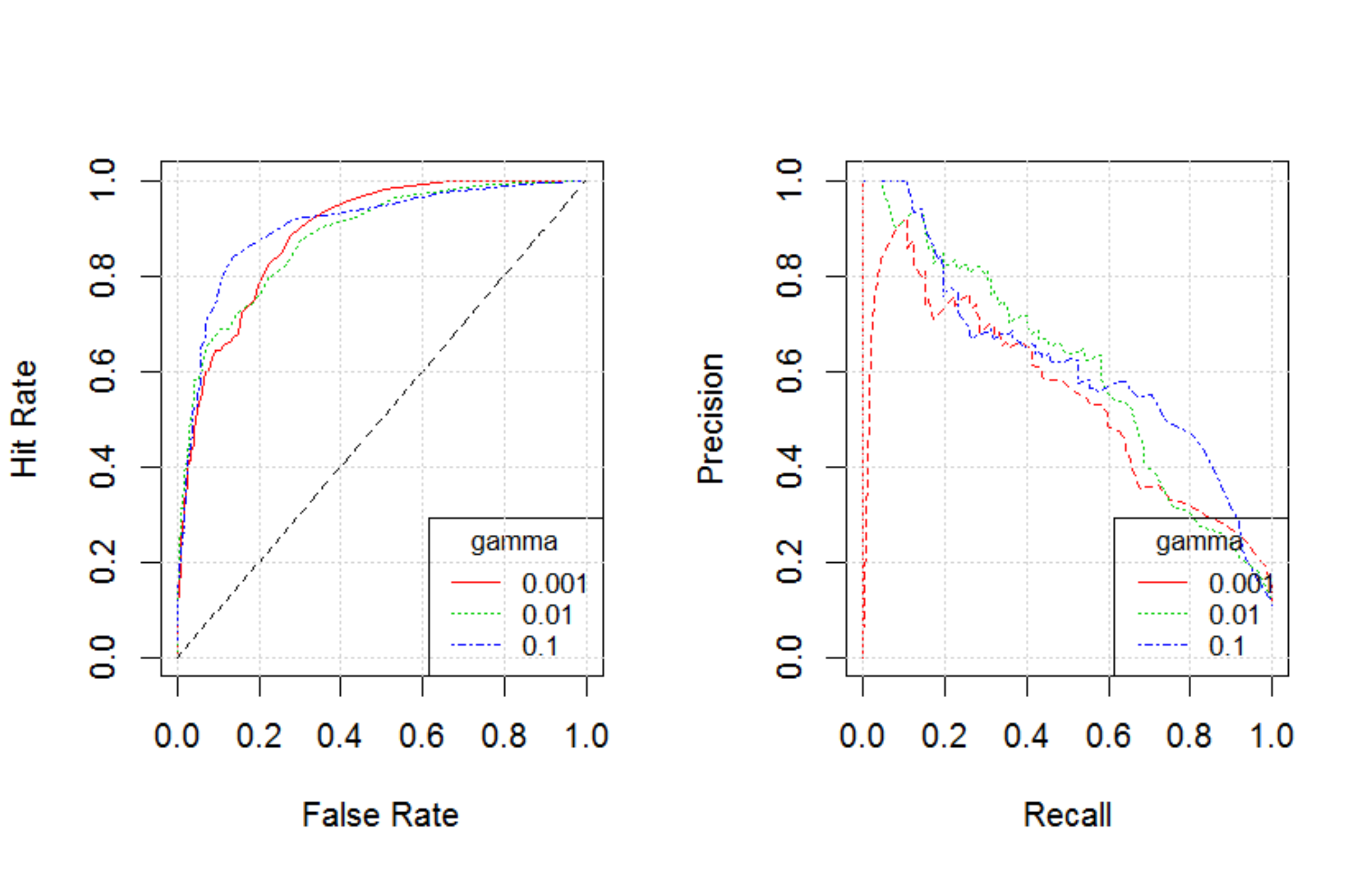}(b)\tabularnewline
\hline 
\hline 
\includegraphics[scale=0.25]{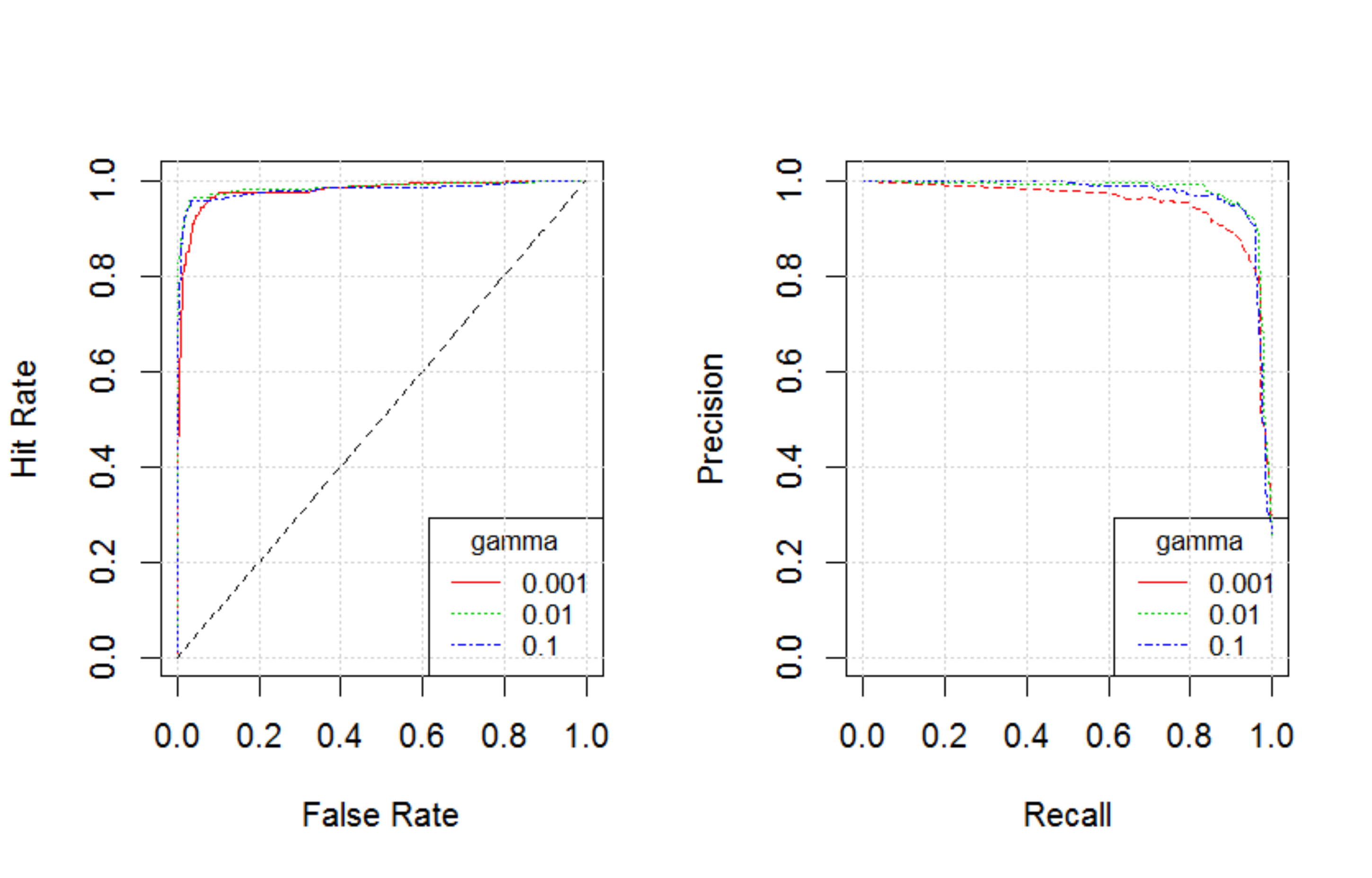}(c) & \includegraphics[scale=0.25]{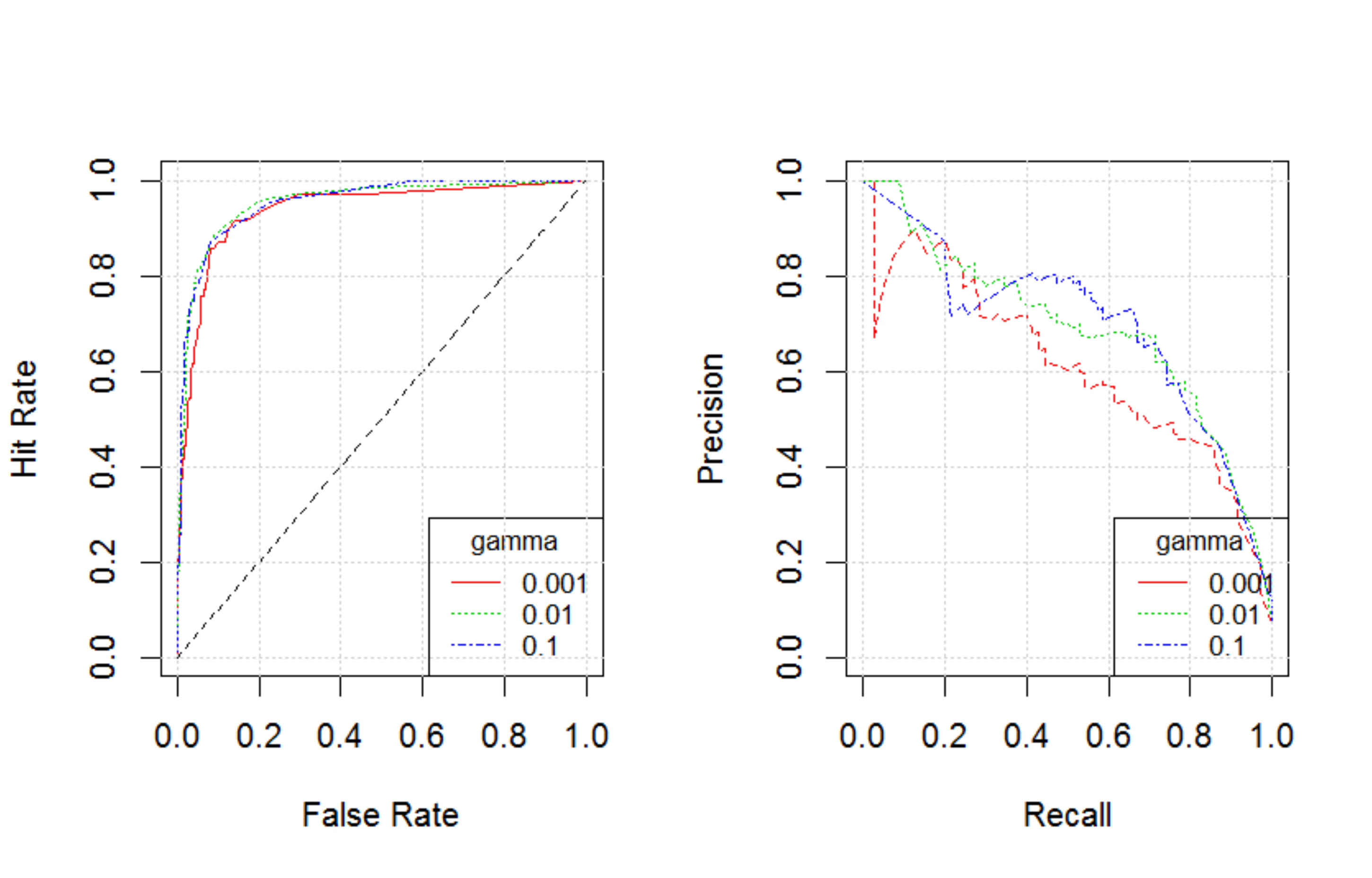}(d)\tabularnewline
\hline 
\end{tabular}
\end{figure}

\begin{figure}[h]

\caption{kNN, (a) Pulsating, (b) Erupting, (c) Multi-Star, (d) Other \label{fig:kNN,-(a)-Pulsating,}}

\begin{tabular}{|c|c|}
\hline 
\includegraphics[scale=0.25]{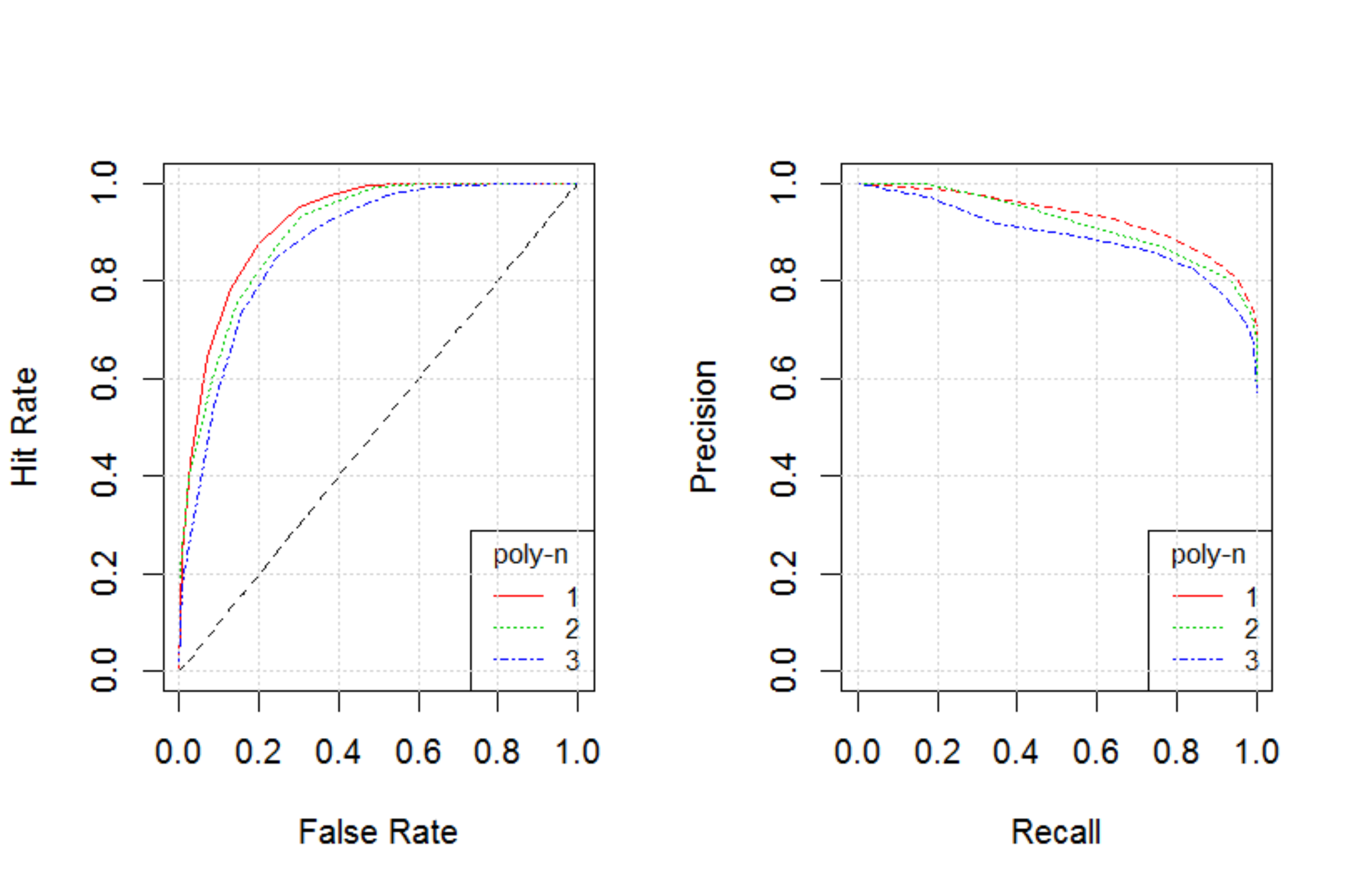}(a) & \includegraphics[scale=0.25]{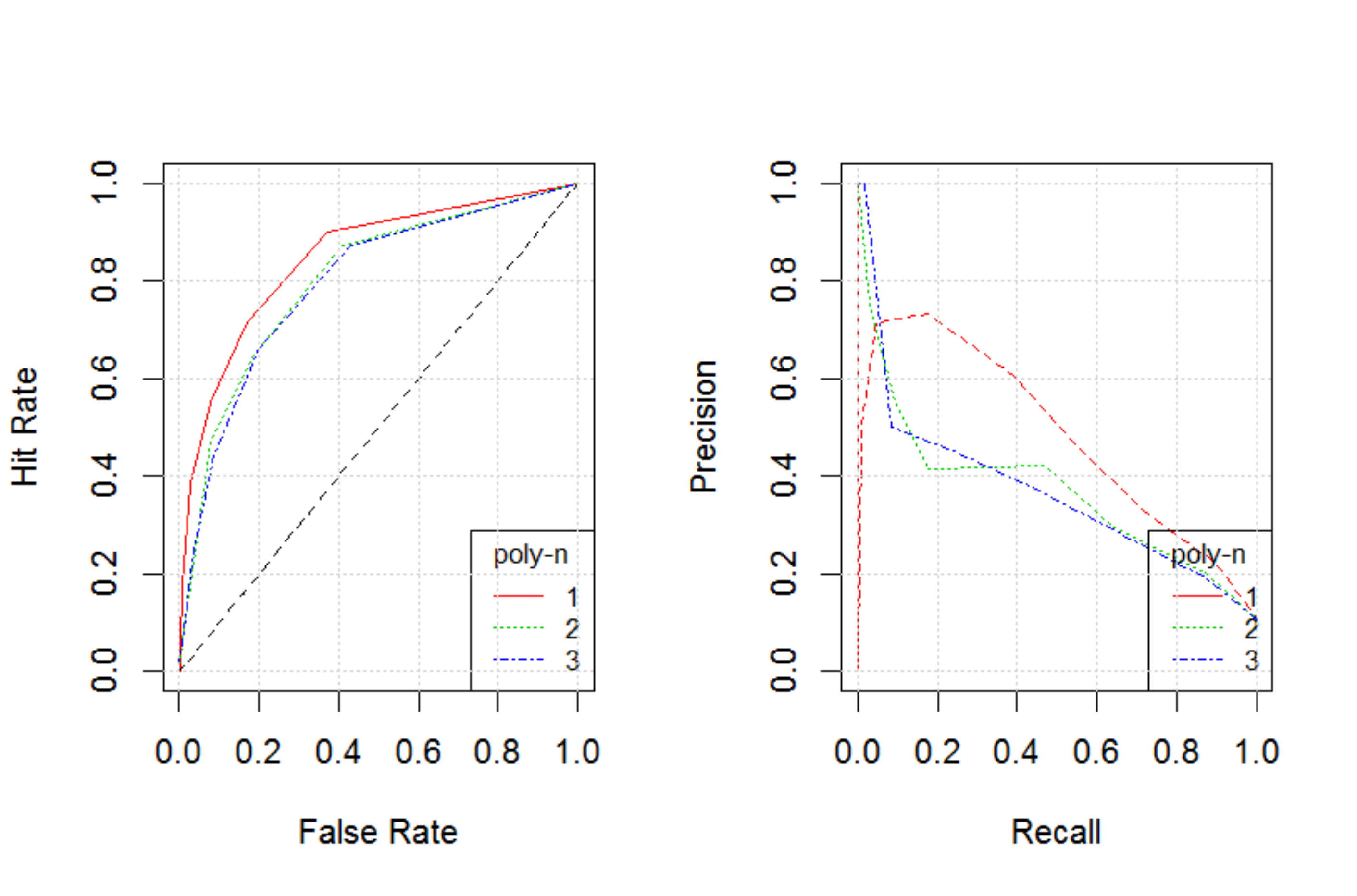}(b)\tabularnewline
\hline 
\hline 
\includegraphics[scale=0.25]{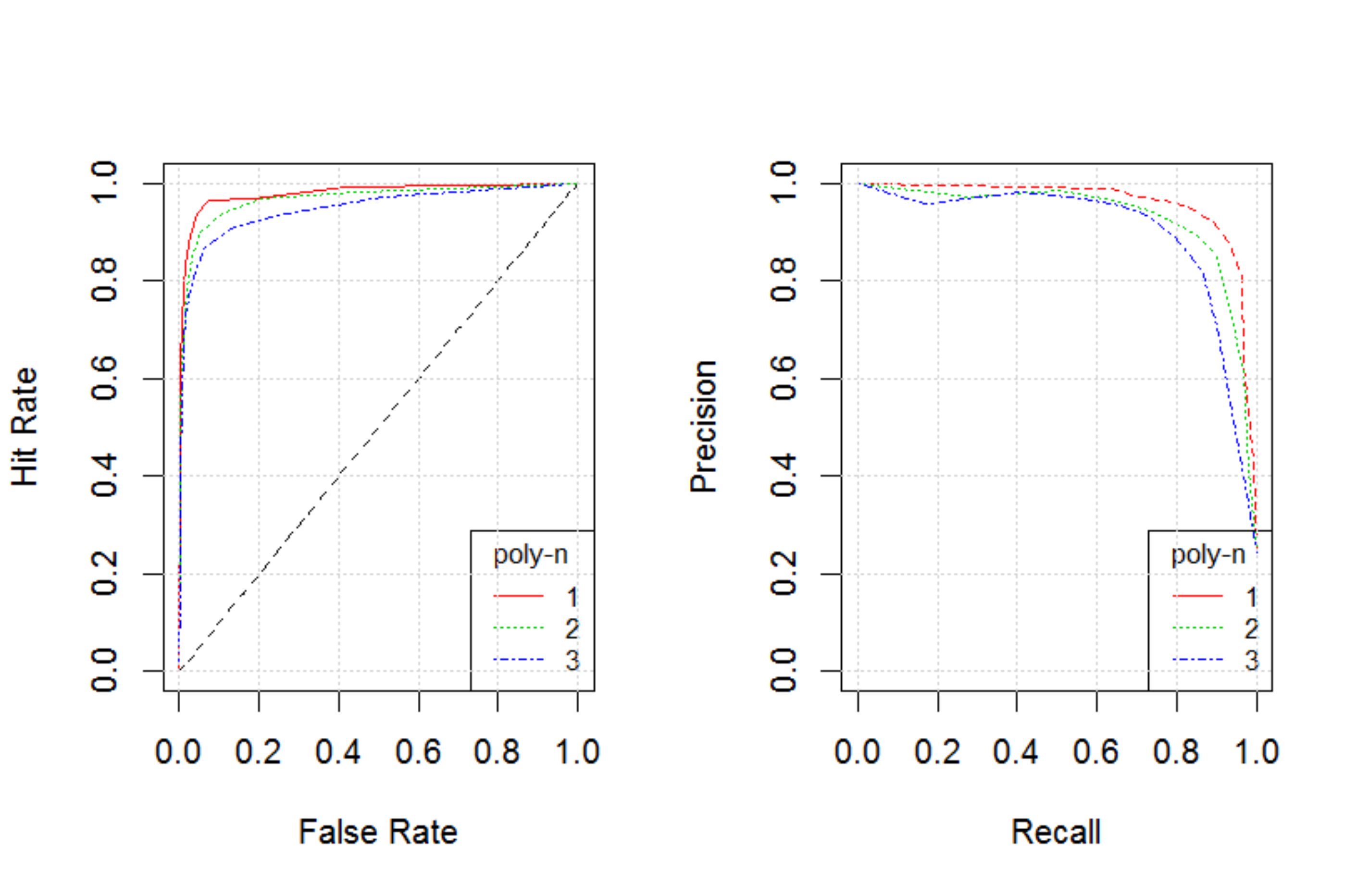}(c) & \includegraphics[scale=0.25]{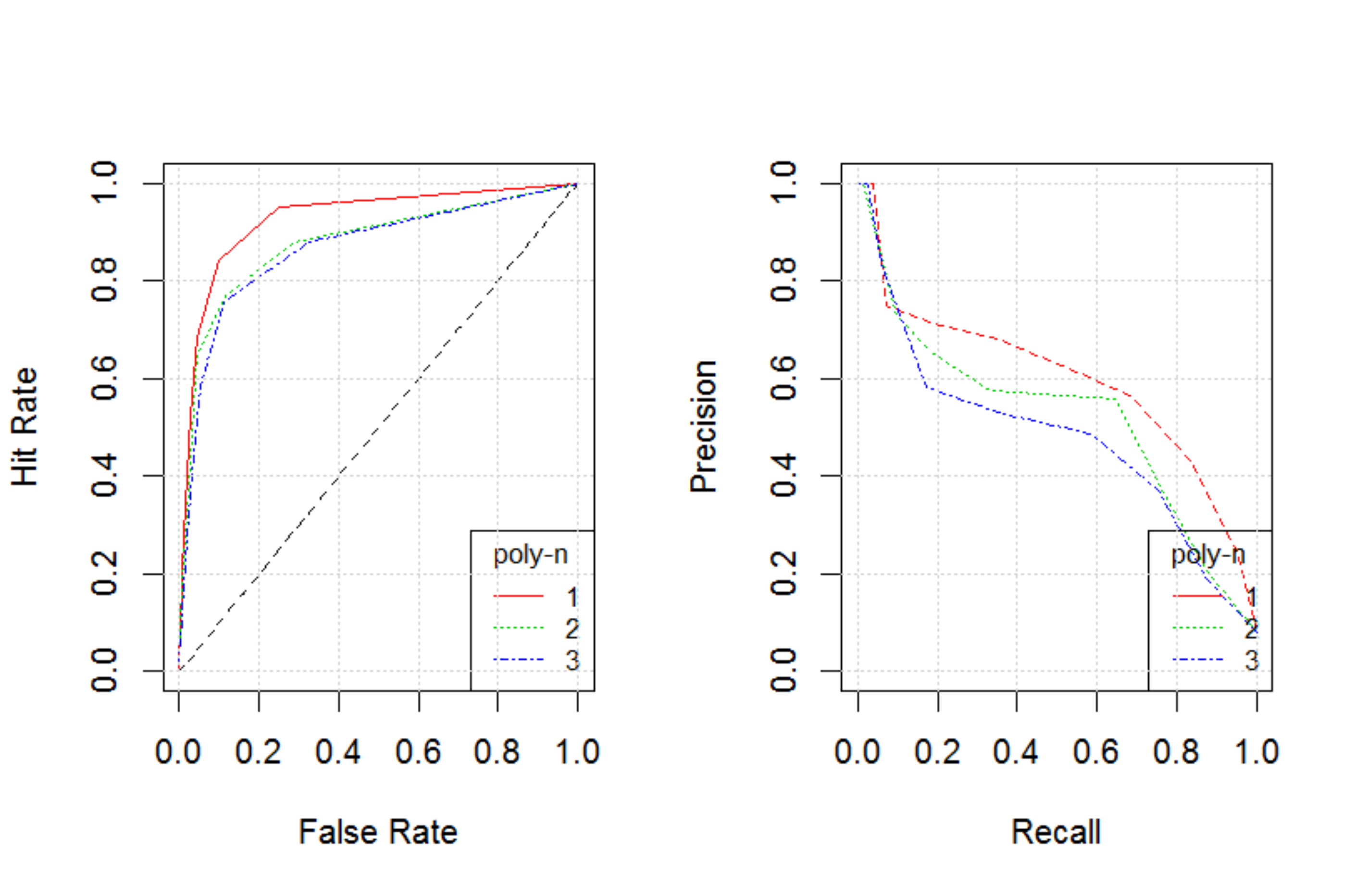}(d)\tabularnewline
\hline 
\end{tabular}
\end{figure}

\begin{figure}[h]

\caption{MLP, (a) Pulsating, (b) Erupting, (c) Multi-Star, (d) Other \label{fig:MLP,-(a)-Pulsating,}}

\begin{tabular}{|c|c|}
\hline 
\includegraphics[scale=0.25]{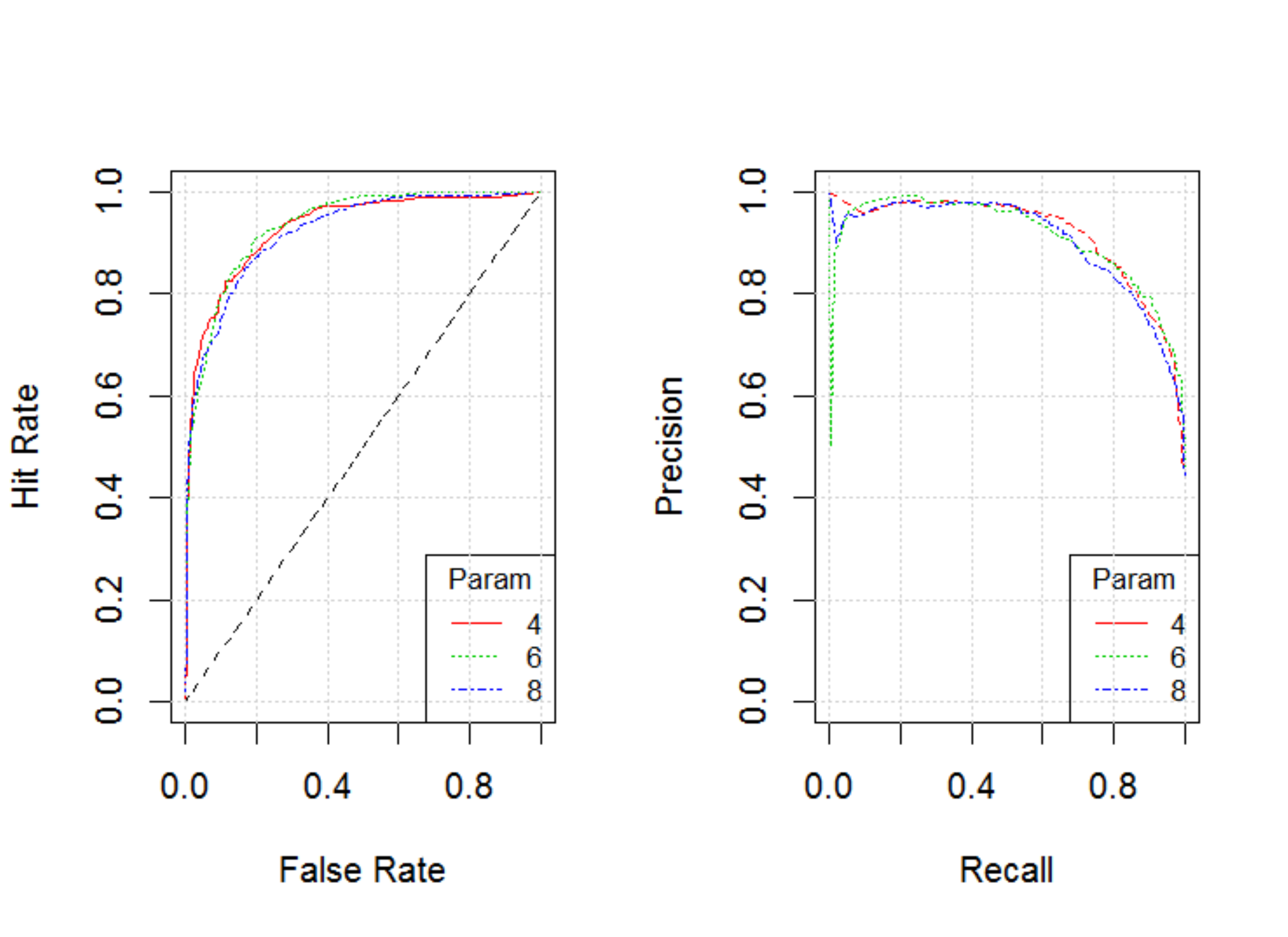}(a) & \includegraphics[scale=0.25]{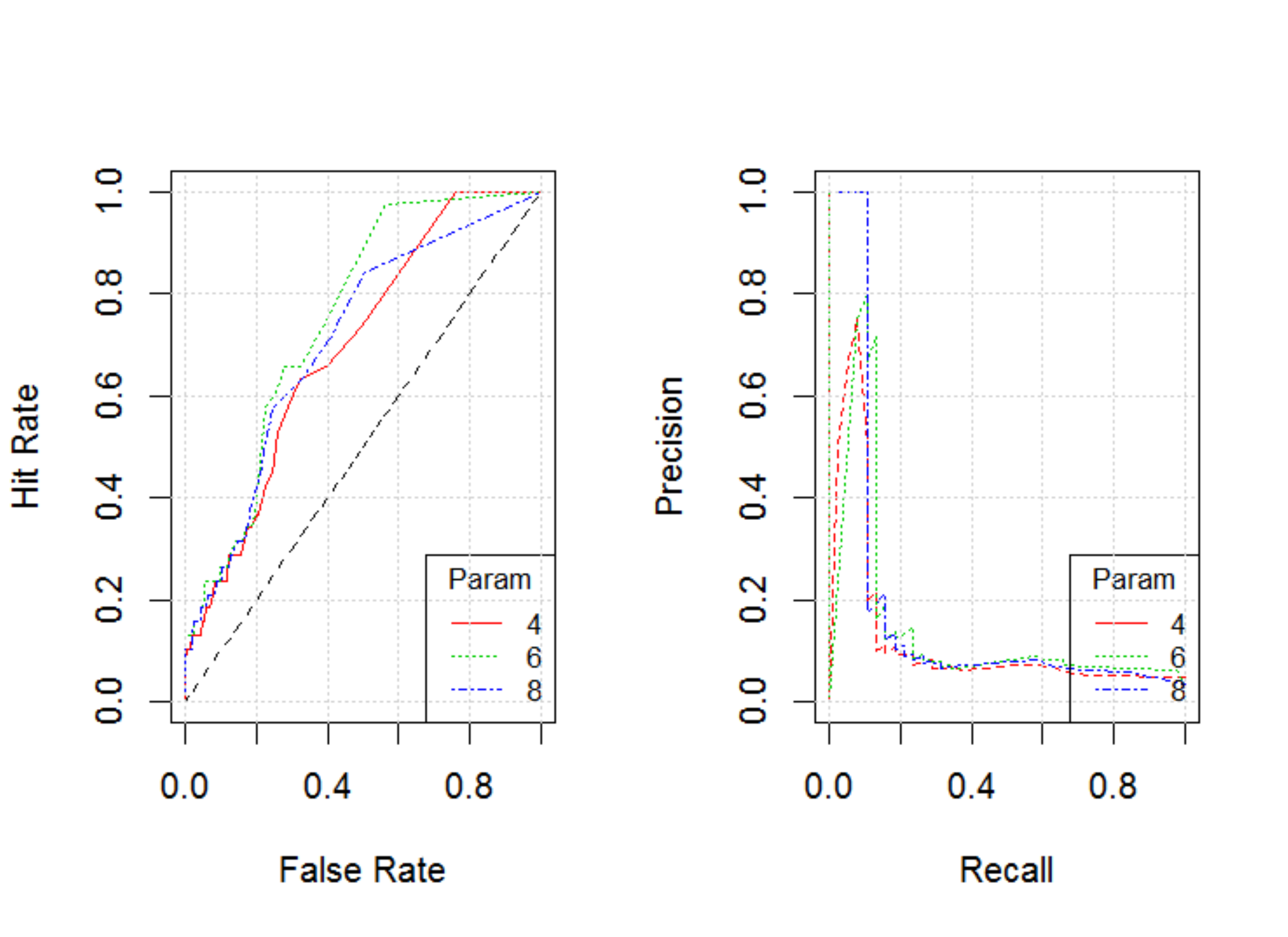}(b)\tabularnewline
\hline 
\hline 
\includegraphics[scale=0.25]{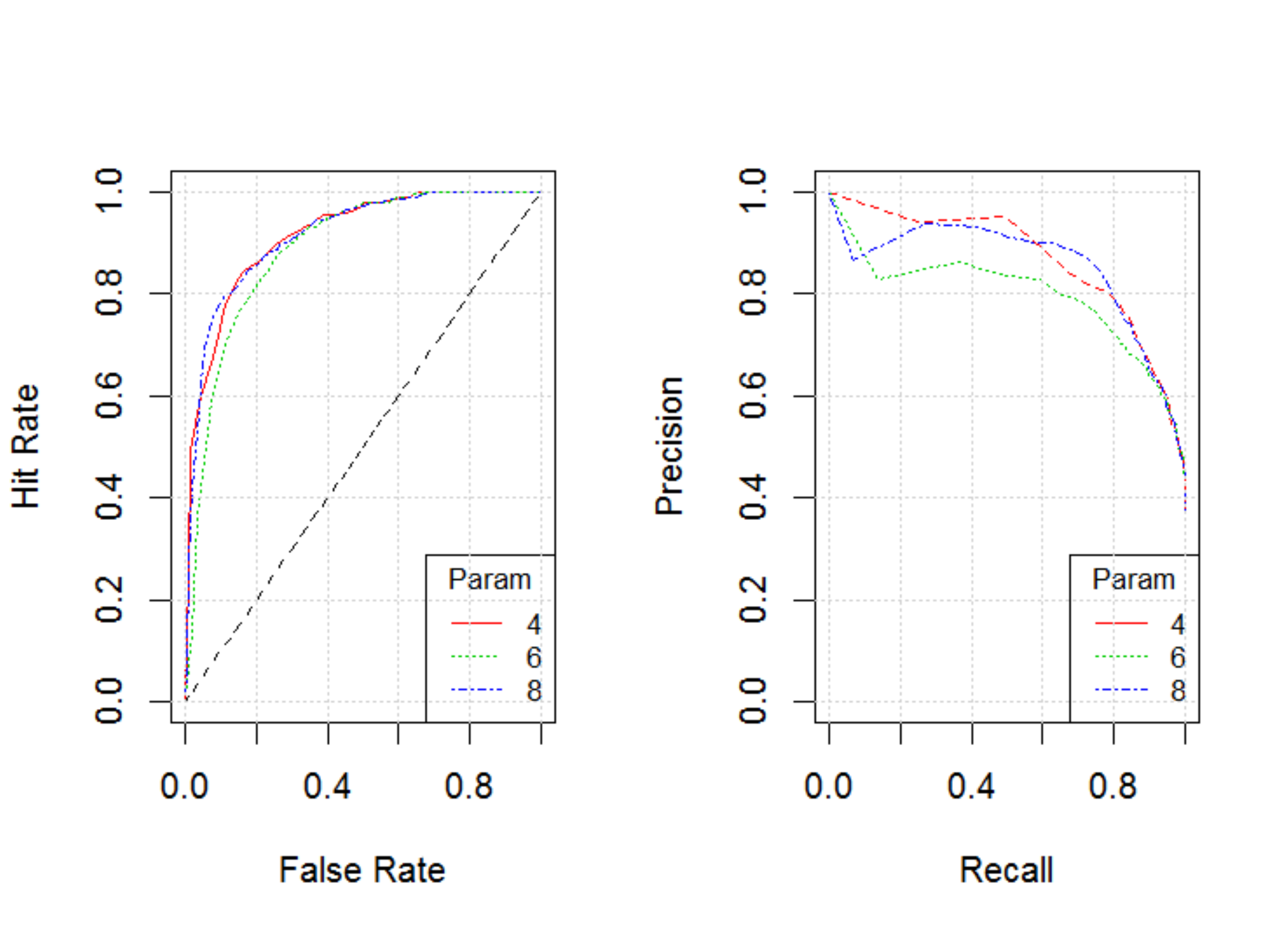}(c) & \includegraphics[scale=0.25]{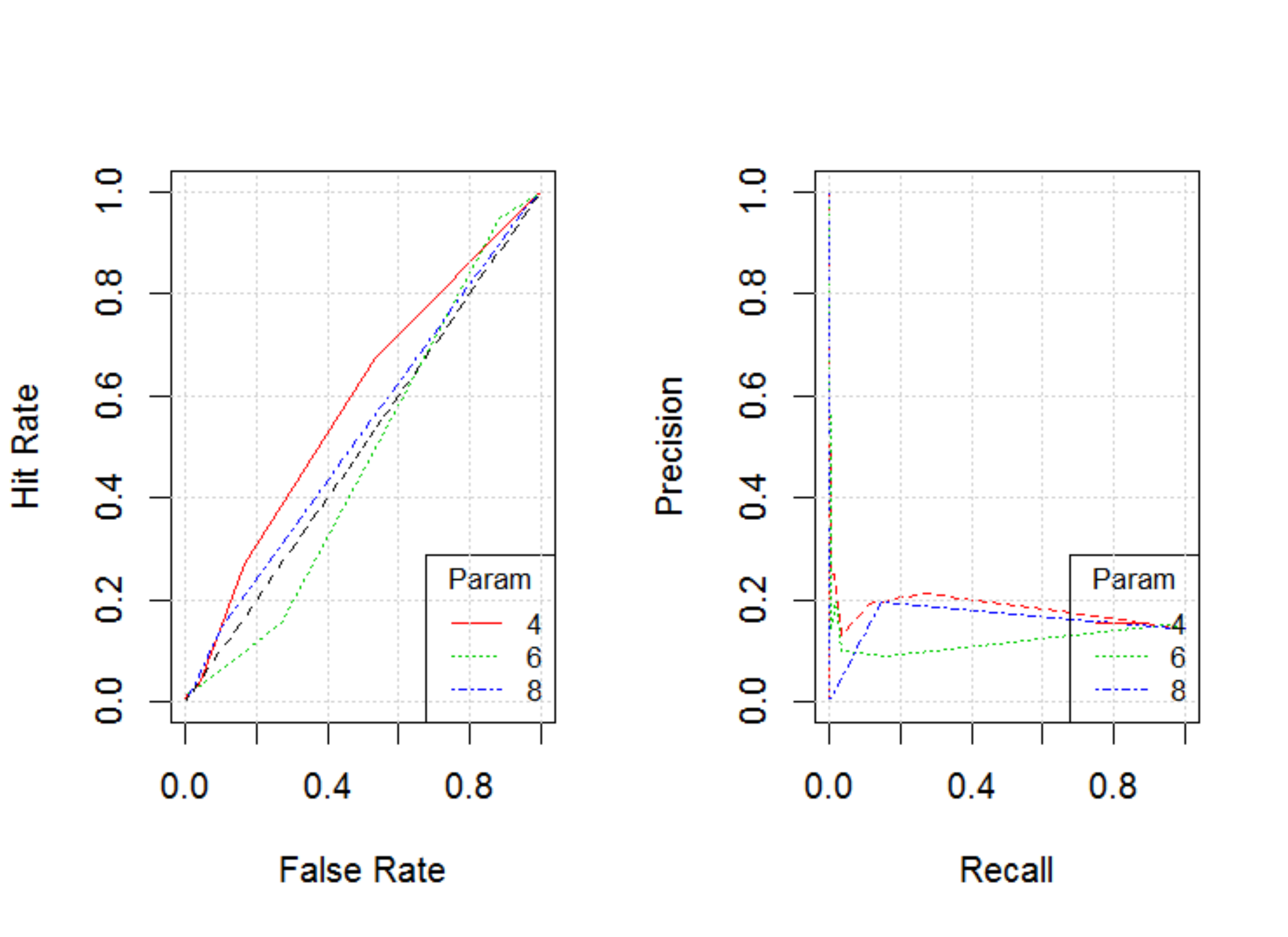}(d)\tabularnewline
\hline 
\end{tabular}
\end{figure}

\begin{figure}[h]

\caption{MLP, Individual Classification, Performance Analysis \label{fig:MLP,-Individual-Classification,}}

\begin{tabular}{|c|c|}
\hline 
\includegraphics[scale=0.25]{ASAS-Hipp-OGLE_MLP_50-50_ROCPR}(a) & \includegraphics[scale=0.25]{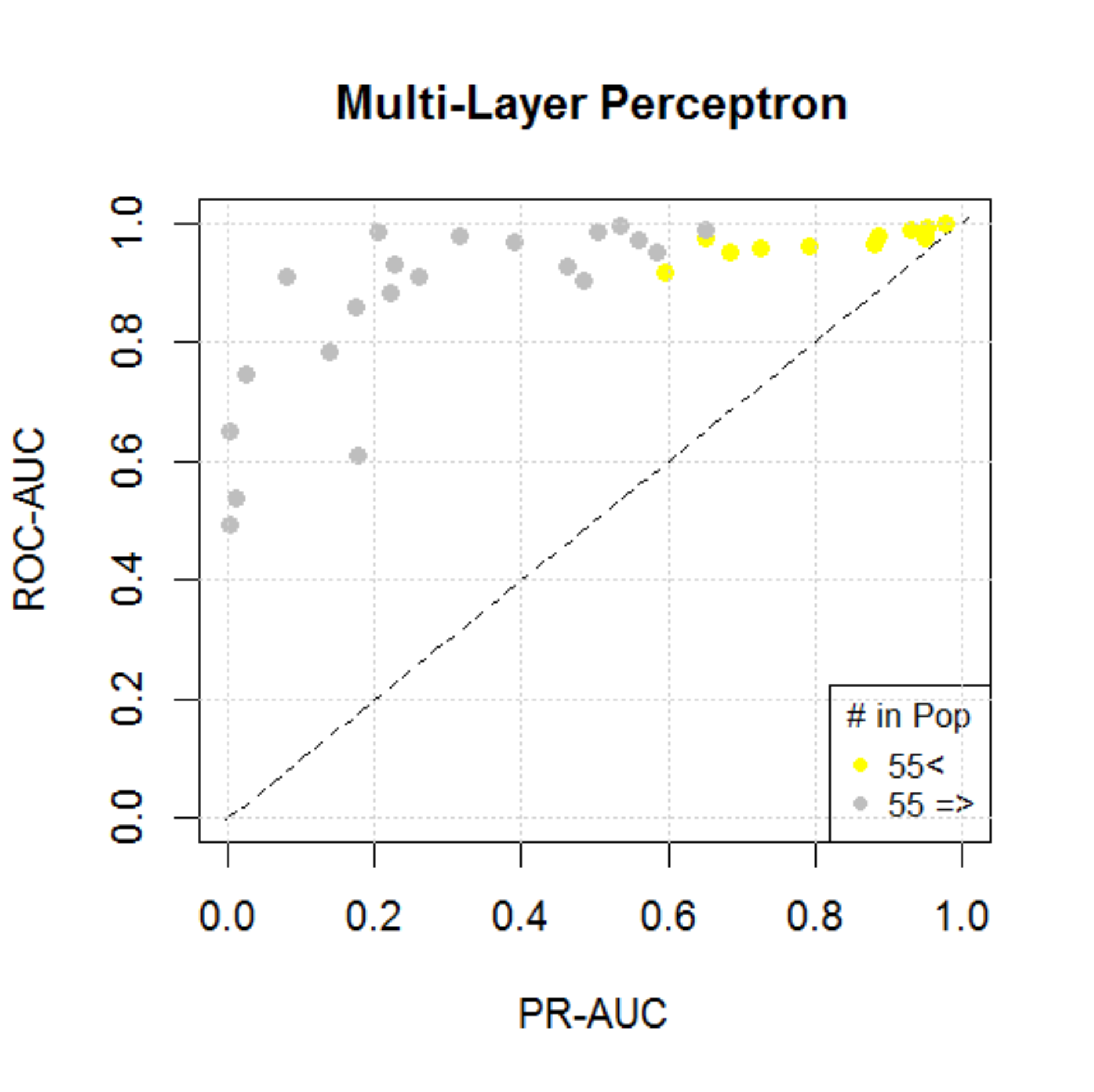}(b)\tabularnewline
\hline 
\end{tabular}
\end{figure}

\begin{figure}[h]

\caption{kNN, Individual Classification, Performance Analysis \label{fig:kNN,-Individual-Classification,}}

\begin{tabular}{|c|c|}
\hline 
\includegraphics[scale=0.25]{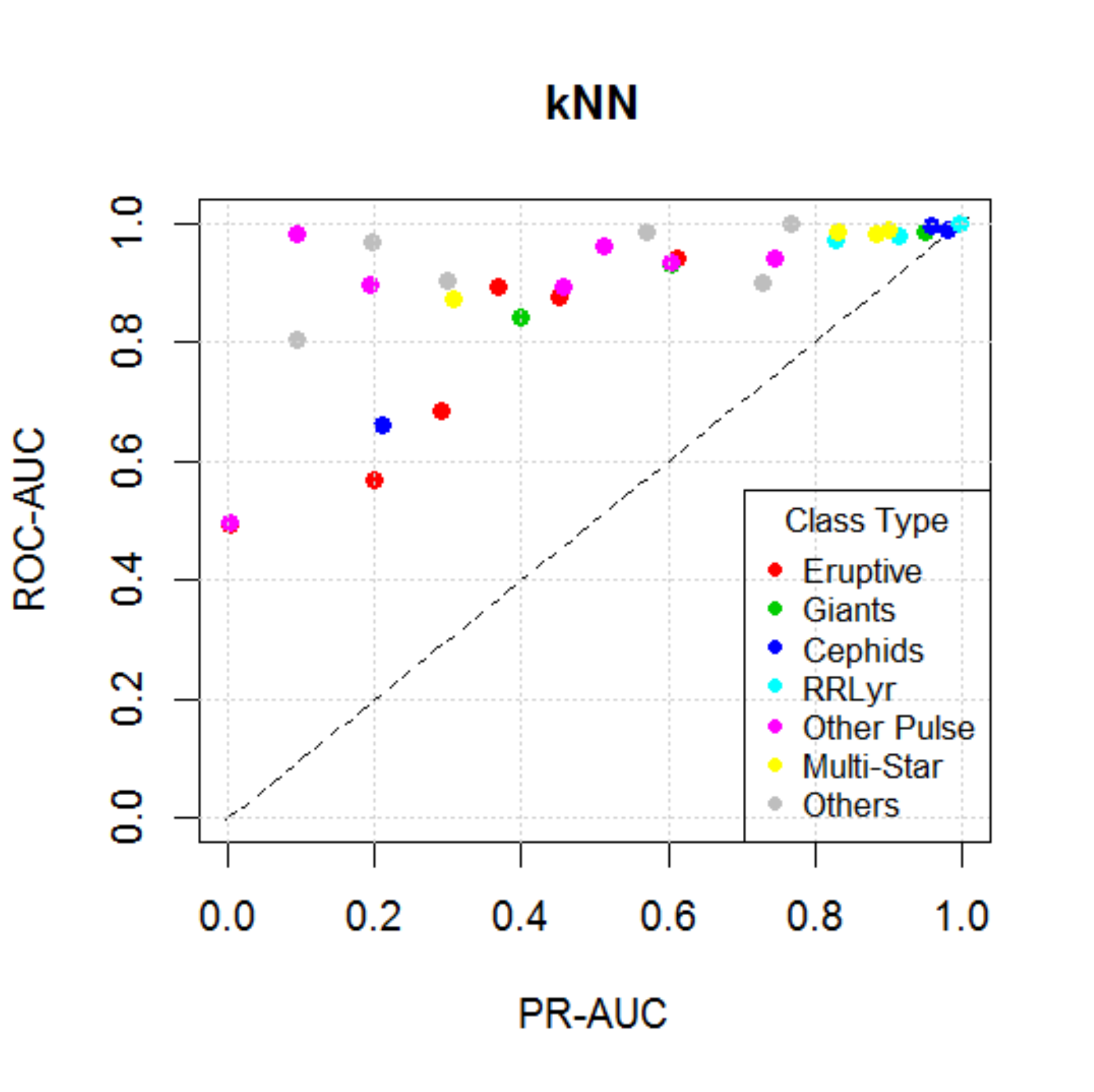}(a) & \includegraphics[scale=0.25]{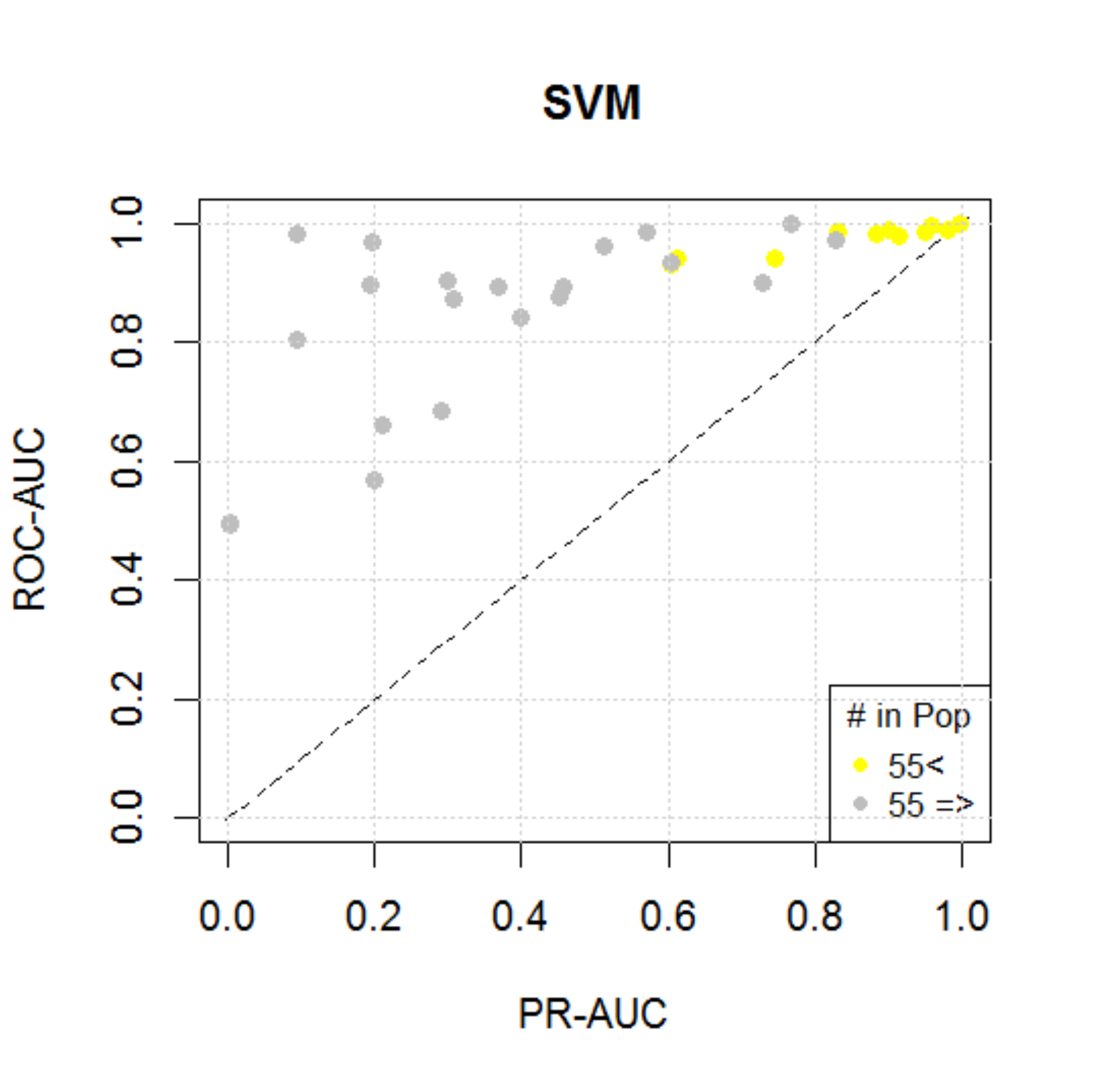}(b)\tabularnewline
\hline 
\end{tabular}
\end{figure}

\begin{figure}[h]

\caption{SVM, Individual Classification, Performance Analysis \label{fig:SVM,-Individual-Classification,}}

\begin{tabular}{|c|c|}
\hline 
\includegraphics[scale=0.25]{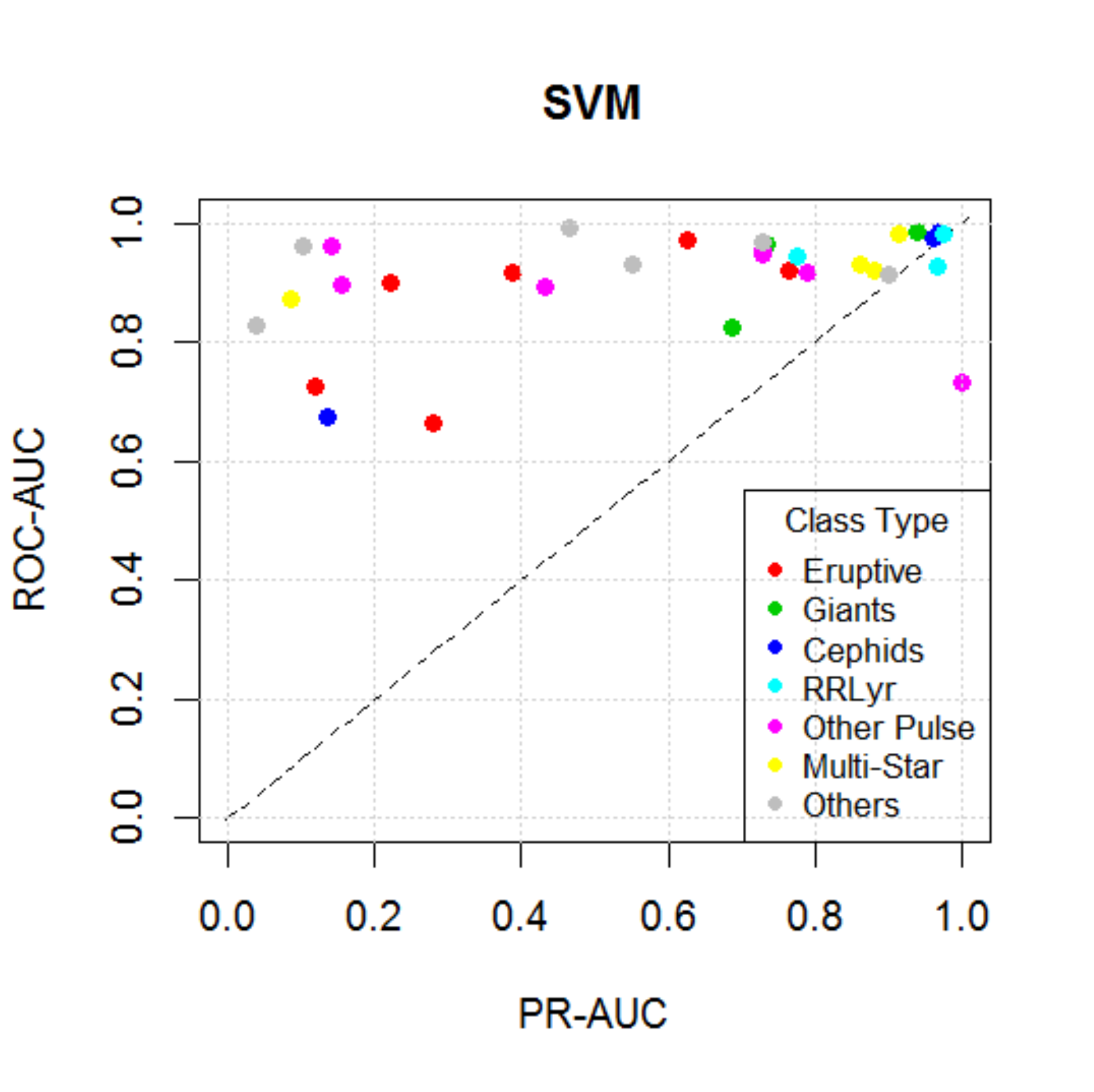}(a) & \includegraphics[scale=0.25]{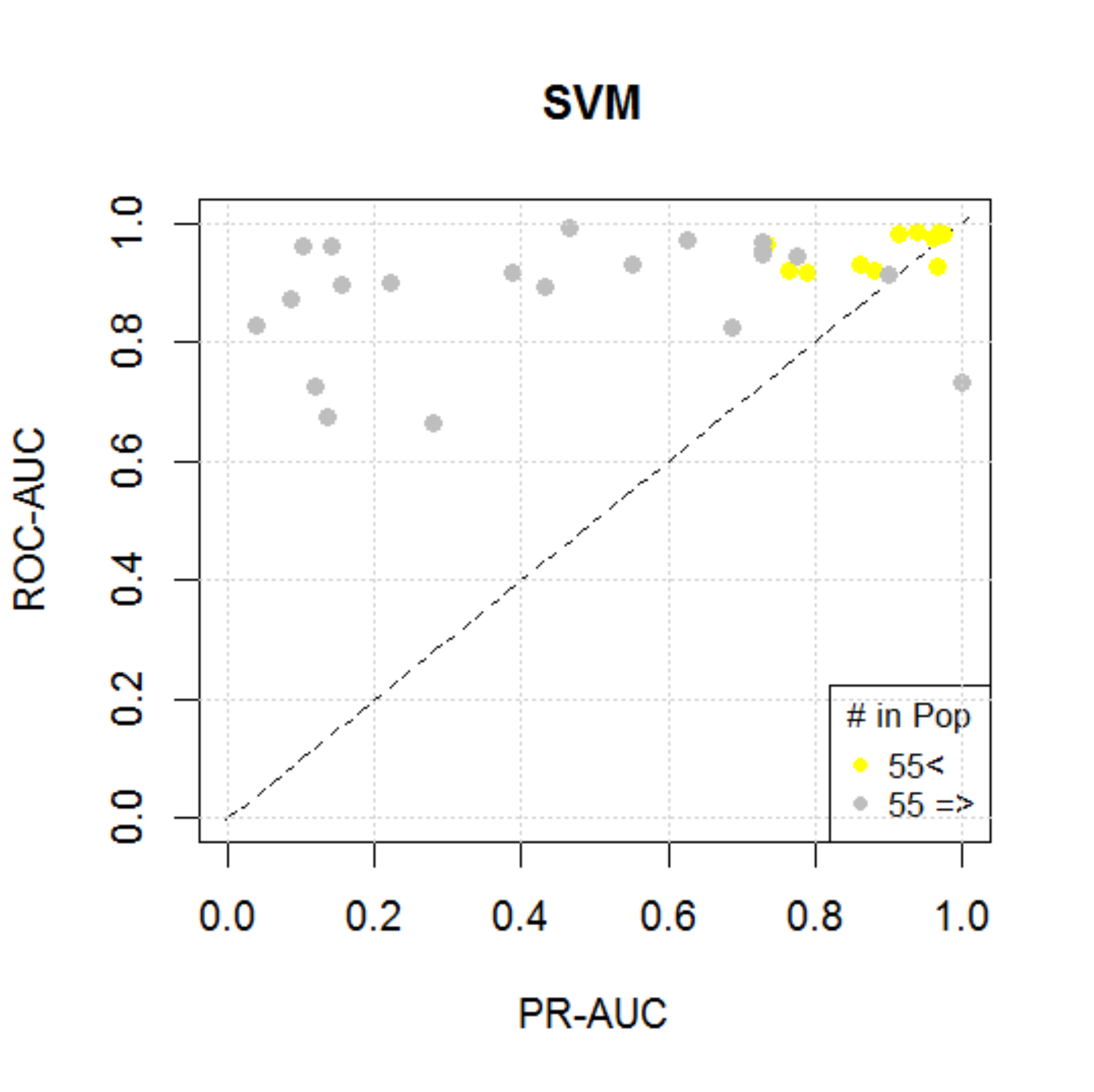}(b)\tabularnewline
\hline 
\end{tabular}

\end{figure}

\end{document}